\documentclass[12pt,letterpaper]{article}
\usepackage[margin=1in]{geometry}
\usepackage[utf8]{inputenc}
\usepackage{amsmath,mleftright,mathtools}
\DeclareMathAlphabet\mathbfcal{OMS}{cmsy}{b}{n}
\usepackage{amsfonts}
\usepackage{amssymb}
\usepackage{graphicx}
\usepackage{float}
\usepackage{mathrsfs}
\usepackage{pdfpages}
\usepackage{physics}
\usepackage{amsthm}
\usepackage{natbib}
\usepackage{multirow}
\usepackage{booktabs}
\usepackage{url}
\usepackage{hyperref}
\usepackage{placeins}
\usepackage{dsfont}
\usepackage{bm}
\usepackage{algorithm}
\usepackage{algpseudocode}
\usepackage{threeparttable}
\usepackage{siunitx}    
\usepackage{setspace}

\usepackage{pifont}

\newtheorem{assumption}{Assumption}

\newtheorem{theorem}{Theorem}
\newtheorem{lemma}{Lemma}
\newtheorem{corollary}{Corollary}
\newtheorem{proposition}{Proposition}

\newcommand{\PX}{P_{X}}
\newcommand{\Xc}{\mathcal{X}}

\newcommand{\E}{\mathds{E}}

\usepackage{mdwlist}

\theoremstyle{definition}
\newtheorem{remark}{Remark}

\usepackage[dvipsnames]{xcolor}

\def \dsE {\text{$\mathds{E}$}}

\DeclareMathOperator{\Var}{Var}

\begin{document}
\setlength{\abovedisplayskip}{0.15cm}
\setlength{\belowdisplayskip}{0.15cm}
\pagestyle{empty}
\newgeometry{top=0.1in,bottom=0.5in,right=1in,left=1in}
\begin{titlepage}
\pagestyle{empty}
\title{
\singlespacing \bf Bayesian Additive Regression Tree Copula Processes for Scalable Distributional Prediction}
\author{Jan Martin Wenkel$^1$, Michael Stanley Smith$^2$ and Nadja Klein$^{1}$\\
   \normalsize $^1$Scientific Computing Center, Karlsruhe Institute of Technology, Germany\\
\normalsize $^2$Melbourne Business School, University of Melbourne, Australia}
\date{}
\maketitle
\vspace{-30pt}
\begin{abstract}
 \singlespacing
We show how to construct the implied copula process of response values from a Bayesian additive regression tree (BART) model with prior on the leaf node variances. This copula process, defined on the covariate space, can be combined with any marginal distribution for the dependent variable to construct a flexible distributional BART model. Bayesian inference is performed via Markov chain Monte Carlo on an augmented posterior, where we show that key sampling steps can be realized as those of \citet{ChiGeoMcC2010}, preserving scalability and computational efficiency despite the high dimensionality of the copula process. The posterior predictive distribution from the copula process model is derived in closed form as the push-forward of the posterior predictive distribution of the underlying BART model with an optimal transport map. Under suitable conditions, and allowing for model misspecification, we show that the model-implied conditional density closest in Kullback-Leibler divergence to the true data-generating process exists, is unique and has a closed form expression. We further establish that fractional posteriors for the conditional densities at the observed covariate values concentrate around this projection at a rate that adapts to the smoothness of the corresponding regression function and to the number of relevant covariates. Simulation studies demonstrate improved accuracy of distributional predictions compared to the original BART model and leading benchmarks. Applications to five real datasets with 506 to 515,345 observations and 8 to 90 covariates further highlight the efficacy and scalability of our proposed BART copula process model.
\end{abstract}
\vspace{-0.5cm}

\singlespacing
\noindent
\mbox{{\bf Keywords}: BART, Distributional Regression, Implicit Copula, KL Projection, Transport Map}\vspace{0.1cm}\\
 \noindent
{\footnotesize{{\bf Corresponding author}: Prof.~Dr.~Nadja Klein, Scientific Computing Center, Karlsruhe Institute of Technology, Zirkel 2, 76131 Karlsruhe, Germany, nadja.klein@kit.edu.\\
\noindent \textbf{Acknowledgments:} Nadja Klein acknowledges support through the Emmy Noether grant KL 3037/1-1 of the German research foundation (DFG). Michael Smith's research has been partially supported by the Australian
Research Council (ARC) Discovery Project grant DP250101069. This work has been supported by the German Federal Ministry of Research, Technology and Space (BMFTR) within the project ``CausalNet'', grant no.~01IS24082}}
\end{titlepage}
\restoregeometry
\newpage
\pagestyle{plain}

\section{Introduction}

Bayesian Additive Regression Trees (BART), introduced by \citet{ChiGeoMcC2010}, is a popular scalable nonparametric approach to regression and classification; see \citet{Hill2020} for a review. BART models the regression function as a sum of regression trees, each constrained by a regularization prior to act as a weak learner~\citep{schapire1990strength}. 
The ensemble of the weak learner trees can capture complex nonlinear relationships and interactions, and 
provides strong predictive performance in high-dimensional and sparse settings; see~\cite{ChiGeoMcC2010,KapBle2016,he2019xbart} for extensive empirical evidence.

A major reason for the success of BART is that it is a well-defined stochastic model that allows for Bayesian uncertainty quantification, along with
fast evaluation of posterior inference and prediction.
It is also easy to extend, and there
is a large literature doing so over the past decade. 
Methodological developments include adaptations to enforce monotonicity constraints \citep{ChiGeoMcCShi2022}, assess sensitivity to violations in modeling assumptions \citep{PraGeoMcC2024}, allow for missing data \citep{LinDan2015}, smoother functions~\citep{LinYan2018} and variable selection and shrinkage \citep{LinDu2023}, including in high-dimensions \citep{Lin2018}. 
Extensions to models for causal inference \citep[][]{HilRic2013,HahMurCar2020}, survival analysis \citep[][]{BasLinSinLip2022,LinBasLiSin2022}, and spatial modeling \citep[][]{MueShiZha2007,GhoSinLinRus2024}, broaden the applicability of BART. In this paper, we consider its extension to distributional regression and prediction using a novel copula process model.

Distributional regression is where the entire distribution of the response variable is modeled as a function of the covariates. It is suitable whenever the response distribution may be heteroscedastic, skewed, heavy-tailed, and/or multi-modal; features which occur in many contemporary applications, see~\cite{Kle2024} for an overview. In this paper, we propose a framework for distributional prediction based on a Gaussian copula process on the covariate space. The copula process is the implicit copula of the response values from a BART model with Gaussian disturbances, and is conditional
on the covariates. We call these ``pseudo-response'' values because they are not observed directly. 
 For the copula process to be effective, we assume a prior on the leaf node variances and derive the distribution of pseudo-response values with the leaf nodes integrated out. 
This approach extends the implicit copulas for regularized linear regression models suggested by~\cite{Klein2019} and~\cite{smithklein21}.
The proposed copula process can then be combined with any marginal distribution for the observed response variable to define a copula process model that is a flexible distributional regression approach that we label ``CP-BART''. Predictions are given by the posterior predictive distribution, which we show is the push-forward of the posterior predictive distribution of the pseudo-response under a transport map that is optimal with respect to quadratic cost. Computation of the predictive distribution and posterior inference use a Markov chain Monte Carlo (MCMC) algorithm that shares key steps with the original sampler of~\cite{ChiGeoMcC2010} and is equally scalable.

{From a theoretical perspective, we do not assume that our CP-BART model is correctly specified and the target of our inference is  the Kullback-Leibler (KL) projection of the true conditional distributions onto the model class implied by the CP-BART. In Lemma~\ref{lem:klproj} we show that this projection exists, is unique and has a closed form expression in terms of the conditional means and variances of the standardized pseudo-response. As a consequence, a hyperprior on the leaf node variances is needed to identify it correctly. Building on the posterior consistency results for BART  \citep{rockova19a,Rokov2020} and the theoretical framework of \citet{Bhattacharya2019}, we then show in Theorem~\ref{thm1} that the {fractional posteriors of our model obtained by raising the likelihood to a fixed power in $(0,1)$ jointly concentrate around the KL projection in the regression function and the leaf-node variances. The resulting contraction rate differs from the minimax rate only by a logarithmic factor that decreases with the number of relevant covariates. A consequence of this which we show in Corollary~\ref{cor1} is that the conditional distributions of the observed responses implied by the model concentrate, in average Hellinger distance, around those of the projection at the same rate. This requires a lower bound on the prior mass of the KL neighborhoods around the projection, which imposes a considerable theoretical challenge treated in Lemma D.2 of the Supplementary Material (SM).}

The finite sample performance of CP-BART is assessed through both simulation studies and
five real world datasets with between 506 and 515,345 observations on 8 to 90 covariates. Density forecasts from CP-BART are more accurate than those from the original BART model of~\cite{ChiGeoMcC2010} and two established benchmark methods for distributional prediction. They are smoother, sharper and/or better located, with a mean pinball (check) loss function that shows they are more accurate at all quantiles in all five real world examples. Following~\cite{LiLinMur2023}, we also show that CP-BART conditional quantile coverage is more accurate, indicating accurate epistemic uncertainty quantification.  

Our paper is not the first to suggest
extending BART to distributional regression. 
\citet{Lin2025} proposed an approach that is applicable to arbitrary parametric distributions assuming the log-density and its derivatives can be provided. While this framework is very flexible, it requires the choice of a set of potential distributions and a parametric assumption that may not hold in practice. A nonparametric alternative is suggested by \citet{OHagRoc2025} who propose a generative regression based on the implicit quantile BART model. This model is different to ours since it targets the conditional quantile function rather than a density forecast. Inference is obtained by a data augmentation approach which can be computationally expensive with increasing sample size. In addition, the authors rely on the  asymmetric Laplace distribution which they frame as a Gibbs posterior \citep{BisHolWal2016} with uncertainty quantification over the quantile function. To avoid quantile crossing in the model by \citet{OHagRoc2025}, the original BART prior needs to be adapted to ensure monotonicity, something that is unnecessary with our approach.  
Our method is also related to other distributional regression models employing BART that do not require a parametric assumption, while modelling the entire distribution.  \citet{OrlMurLinVol2021} consider a covariate-dependent mixture model for a latent variable. This work is related to the adaptive conditional density estimation model of  \citet{LiLinMur2023}, who model the conditional density of the response by tilting a base model and modelling the regression function via the soft decision tree of \citet{LinYan2018}. While not the focus of this paper, it is conceptually also possible to extend our approach to multivariate responses. An example of such a multivariate response model with BART is that of   \citet{UmLinSinBan2023}, who consider the multivariate skew normal distribution with a multivariate BART prior. 

The rest of the paper is structured as follows. Sections \ref{sec:model} and \ref{sec:estimation} detail the proposed model and efficient Bayesian estimation, respectively. In Section \ref{sec:theory} we establish posterior consistency.  Sections \ref{sec:simulations} and \ref{sec:real}  evaluate our CP-BART model empirically in simulations and real world data illustrations. Section~\ref{sec:discussion} concludes by highlighting the contributions of our work and directions for further work.

\section{Proposed Model}\label{sec:model}
In this section we first briefly specify the BART model, introducing the necessary notation for our theoretical results in Section \ref{sec:theory}, and the priors that we adopt. We then write it as a Gaussian linear model, which is used to construct the implicit
copula process of its response values. Finally, we show how to construct the posterior predictive distribution
from the copula process model with arbitrary marginals using a transport map. This is the basis of our proposed distributional BART model.

\subsection{Bayesian Additive Regression Trees}\label{sec:BART}
BART was introduced by \cite{ChiGeoMcC2010} and is an extension of classical sum of trees models in which the unknown regression function is modelled as a flexible ensemble of trees.

\paragraph{Regression Tree:}
A single tree with partition $\mathcal{T}=\{T_k\}_{k=1}^K$ of the unit cube $[0,1]^p$ for $n$ realizations $x_1,\dots,x_n$ of a $p$-dimensional independent variable vector (standardized to the unit cube), has empirical measure
\begin{equation*}
    \mu(T_k) = \frac{1}{n}\sum_{i=1}^n\mathds{1}_{T_k}(x_i)\,,
\end{equation*}
where the indicator function $\mathds{1}_A(x)=1$ if $x\in A$ and zero otherwise.  A tree partition is called valid if it satisfies
\begin{equation*}
    \mu(T_k)\geq \frac{C_0^2}{n}\quad\,,\hspace{5pt}\mbox{ for } k = 1,\dots,K,
\end{equation*}
for some constant $C_0^2\in\mathds{N}\setminus\{0\}$. Now consider the function space containing all possible step functions from valid tree partitions defined as 
\begin{equation*}
    \mathcal{F}_\mathcal{T} = \bigcup_{q=0}^\infty\bigcup_{K=1}^\infty\bigcup_{Q:|Q|=q}\mathcal{F}(\mathcal{V}_Q^K)\,,
\end{equation*}
where $\mathcal{V}_Q^K$ denotes the family of valid tree partitions with $K$ cells splitting each $x_i$ at least once. Here, $Q\subseteq\lbrace 1,\ldots,p\rbrace$ with $|Q| = q \leq p$ denotes the number of ``active'' covariates that determine the splits. Thus, $\mathcal{F}_\mathcal{T}$ is the union over all possible numbers of active predictors, tree sizes and partitions.

For a tree partition $\mathcal{T}\in\mathcal{V}_Q^K$ with leaf node values denoted as $M=(\mu_1.\ldots,\mu_K)^\top \in \mathbb{R}^K$, the associated set of tree-structured step functions is defined by 
\begin{equation*}
\mathcal{F}(\mathcal{V}_Q^K) = \left\{f_{\mathcal{T},M}:[0,1]^p \rightarrow \mathbb{R};\ f_{\mathcal{T},M}(x) = \sum_{k=1}^K\mu_k\mathds{1}_{T_k}(x), \mbox{ where } \mathcal{T}\in\mathcal{V}_Q^K,\,M \in \mathbb{R}^K  \right\}.    
\end{equation*}

\paragraph{Additive Regression Tree:} An additive regression tree model extends a single tree by combining $m$ trees into an ensemble. For tree $j$, 
let $Q^j\subseteq\lbrace 1,\ldots,p\rbrace$ be its $|Q^j| = q^j$ active covariate indices, and $K^j$ be the number of terminal nodes. 
If $f_{\mathcal{T}^j,M^j}\in\mathcal{V}_{Q^j}^{K^j}$ denotes the step function associated
with tree $j$, a regression forest can be defined as the ensemble
\begin{equation*}
    f_{\mathcal{E},\mathcal{M}}(x)=\sum_{i=1}^mf_{\mathcal{T}^j,M^j}(x),
\end{equation*}
where $\mathcal{E}=\{\mathcal{T}^1,\dots,\mathcal{T}^m\}$ is the set of trees in the ensemble and $\mathcal{M}=(M^1,\dots,M^m)^\top$ is the vector of all the leaf nodes in the ensemble. 

Moreover, if the covariate indices $\mathcal{Q} = \{Q^1,\dots, Q^m\}$ and tree sizes $\mathcal{K} = (K^1,\dots,K^m)^T$, we use $\mathcal{VE}_\mathcal{Q}^\mathcal{K}$ to denote the ensemble of tree partitions $\mathcal{T}^j$, $j=1,\ldots,m$.   
We can then define the function space $\mathcal{F}(\mathcal{VE}_\mathcal{Q}^\mathcal{K})$ as
\begin{equation*}
\mathcal{F}(\mathcal{VE}_\mathcal{Q}^\mathcal{K}) = \left\{f_{\mathcal{E},\mathcal{M}}:[0,1]^p \rightarrow \mathbb{R};\ f_{\mathcal{E},\mathcal{M}}(x) = \sum_{j=1}^mf_{\mathcal{T}^j,M^j}(x); \mathcal{E}\in\mathcal{VE}_\mathcal{Q}^\mathcal{K}, M^j \in \mathbb{R}^{K^j}  \right\}.
\end{equation*}

\paragraph{Prior distributions:}
A defining feature of BART is that the prior regularizes each tree so that it behaves as a ``weak learner''~\citep{schapire1990strength,ChiGeoMcC2010} where no single tree explains much of the variation, and only the sum produces a flexible model. This is achieved through hierarchical priors on the structure and leaf values of each tree.
 For the number of active predictors $q^j = |{Q}^j|$ in each tree $j$, we assume
 the prior
\begin{equation*}
\pi(q^j) \propto C^{-q^j}p^{Aq^j}\,,\hspace{5pt}\mbox{ for } q^j = 0,1,\dots,p,
\end{equation*}
and some constants $A,C>0$. Conditional on $q_j$, a uniform prior 
on $Q^j$ is assumed, so that
\begin{equation*}
    \pi({Q}^j\ |\ q^j) = \binom{p}{q^j}^{-1}.
\end{equation*}

Following~\cite{rockova19a}, the
prior on the number of terminal nodes $K^j$ follows a Galton–Watson branching process with 
node-specific splitting probability $p_{\textit{split}}(k) = \nu^{d(k)}$ 
where $d(k)$ denotes the depth of node $k$ in the tree, 
 $0<\nu\leq 1$.  
 Given $\mathcal{Q}$ and $\mathcal{K}$, a (conditionally) uniform prior is placed on the tree topologies $\mathcal{E}$:
\begin{equation*}
    \pi(\mathcal{E}\ |\ \mathcal{Q}, \mathcal{K}) = \frac{1}{|\mathcal{VE}_\mathcal{Q}^\mathcal{K}|}\mathds{1}_{\mathcal{VE}_\mathcal{Q}^\mathcal{K}}(\mathcal{E}).
\end{equation*}
Finally, independent zero mean Gaussian priors on the leaf node values of each are assumed. We also do so when constructing the copula process, but where the variance is selected differently than in~\cite{ChiGeoMcC2010}, as discussed further below.

\begin{remark}\
The prior on $K^j$ differs from that originally adopted by~\cite{ChiGeoMcC2010}. \cite{rockova19a}
argue that the original prior provides insufficient penalization of large of trees to achieve the necessary tail decay required to show posterior consistency using their proof strategy. They propose what they call the Galton-Watson prior which is the one we adopt for our theoretical results in Section~\ref{sec:theory}.  
Moreover, we found both the original and adopted priors give very similar results empirically.
\end{remark}

\subsection{BART Copula Process Model}\label{sec:BARTCPM}
With Gaussian disturbances, we write BART as a Bayesian linear model from which its implicit copula process is then constructed. We then show that the copula process is interpretable as a transport map, which then allows for efficient Bayesian inference in Section~\ref{sec:estimation}.
When the copula process is combined with an arbitrary marginal for a dependent variable, the resulting copula process model is shown to provide flexible distributional predictions.

\paragraph{BART Linear Model:}
Consider the following regression for a response $\widetilde{Z}_i$ (which we call the ``pseudo-response'') with independent additive Gaussian disturbances
\begin{equation}\label{eq:regression}
    \widetilde{Z}_i = f_0(x_i) + \varepsilon_i,\quad\varepsilon_i\sim\mathcal{N}(0,\sigma^2)\,,
\end{equation}
for $i=1,2,\ldots,n$. The mean function $f_0$ is unknown, and we model it using BART as follows.
Recall that $M^j$ is the vector of leaf node values of the ensemble member $j$. Then let $E^j$ be a $(n\times K^j)$ matrix of zeroes and ones that selects the elements of $M^j$, such that
\begin{equation}
   {\rm f}_{\mathcal{T}^j,M^j} := \left(f_{\mathcal{T}^j,M^j}(x_1),\ldots,f_{\mathcal{T}^j,M^j}(x_n)\right)^\top = E^jM^j.
\end{equation}
If $E = \left[E^1\ |\ E^2\ |\ \cdots\ | \ E^m\right]$, then the ensemble evaluated at the $n$ observations can be written as
\begin{equation*}
\left(f_0(x_1),\dots,f_0(x_n)\right)^\top=\left(f_{\mathcal{E},\mathcal{M}}(x_1),\ldots,f_{\mathcal{E},\mathcal{M}}(x_n)\right)^\top = \sum_{j=1}^m {\rm f}_{\mathcal{T}^j,M^j} = E\mathcal{M}\,. 
\end{equation*} 
Conditional on  $\{\mathcal{E},\mathcal{M},\sigma^2\}$, the pseudo-response vector
 $\widetilde{Z}_{1:n}=(\widetilde{Z}_1,\ldots,\widetilde{Z}_n)^\top$ follows the linear model
\begin{equation}\label{eq:modz}
    \widetilde{Z}_{1:n} | \mathcal{E},\mathcal{M},\sigma^2 \sim \mathcal{N}(E\mathcal{M},\sigma^2I_n).
\end{equation}
where  $I_n$ is the $n\times n$ identity matrix, and the trees $\mathcal{E}$ summarize all the 
information in the covariates $x_{1:n}=\{x_1,\ldots,x_n\}$.
We adopt the BART prior as outlined previously. For the leaf node values we choose independent Gaussian priors of the form 
\begin{equation}\label{eq:Mprior}
M^j\mid c,\sigma^2 \sim\ \mathcal{N}(0,c\sigma^2 I_{K^j})\,,\hspace{5pt}\mbox{ for }  j=1,\ldots,m\,,
\end{equation}
where $c>0$.
The prior variance differs from that adopted by~\cite{ChiGeoMcC2010} in two ways. First, scaling by $\sigma^2$ simplifies its integration out of the copula process, where later it is shown to be unidentified. Second, we estimate the scale parameter $c$,
for which we adopt the inverse gamma prior $c\sim\mathcal{IG}(a,b)$, with shape and scale parameters $a,b$. We show later posterior consistency under this prior, how to select $a,b$, and that this additional layer of flexibility improves predictive accuracy of the copula model. We demonstrate this empirically in Section \ref{sec:simulations} and find in addition, that a prior on $c$ for the standard BART model does not yield similar benefits. 

\paragraph{Gaussian Copula Process:}
Every multivariate continuous distribution has a unique copula that fully captures its dependence structure, widely called its ``implicit copula''~\citep{SmithES2023}. We employ the implicit copula 
of the conditional distribution of $\widetilde{Z}_{1:n}$ with the leaf node values marginalized out given by
\begin{equation}\label{eq:condgaussian}
\begin{split}
    \widetilde{Z}_{1:n}|\mathcal{E},c,\sigma^2\ & \sim 
    \mathcal{N}\left(0,\sigma^2(cEE^T+I_n)\right)\,.        
\end{split}
\end{equation}
The implicit copula of a Gaussian distribution is the popular Gaussian copula and its  parameter is the correlation matrix of the distribution; see~\cite{song2000}. Some simple computations show that the correlation matrix of~\eqref{eq:condgaussian} is $\Omega=S_{1:n}(cEE^\top+I_n)S_{1:n}$, where the scaling matrix $S_{1:n} = sI_n$ with $s =\mbox{var}(\tilde{Z}_i)^{-1/2}= (1+mc)^{-\frac{1}{2}}$. 

Let $u_{1:n}\in [0,1]^n$ and $z_{1:n}=\underline{\Phi}^{-1}(u_{1:n})$, where $\underline{\Phi}^{-1}$ denotes elementwise application of the standard normal quantile function. Then the density of the Gaussian copula process is 
\begin{equation}
c_{\mbox{\tiny Ga}}(u_{1:n};\Omega(c,\mathcal{E}))=\left|\Omega(c,\mathcal{E})\right|^{-1/2}
\exp\left(-\frac{1}{2}z_{1:n}^\top\left(\Omega(c,\mathcal{E})^{-1}-I_n\right)z_{1:n}\right)\,, \label{eq:gcopprocess}
\end{equation}
and we make three observations. First, $\Omega$ is not a function of $\sigma^2$, which shows that the scale of $\widetilde{Z}_i$ is unidentified in the copula. Therefore, without loss of generality, we set $\sigma=1$ in~\eqref{eq:regression} throughout.
Second,
$\Omega$ is a function of $c,\mathcal{E}$, which are the only parameters of our Gaussian copula process. Therefore, we sometimes write $\Omega(c,\mathcal{E})$ for $\Omega$, and later show how to use their Bayesian posterior for inference on their values. Third, marginalizing 
the leaf node values $\mathcal{M}$ out of~\eqref{eq:condgaussian} is necessary when constructing the copula process. If it was not, then the implicit copula of $\widetilde{Z}_{1:n}|\mathcal{E},c,\sigma^2,\mathcal{M}$ at~\eqref{eq:modz} is simply the independence copula, which provides uninformative predictions of future observations.

\paragraph{Prior Scaling:}
Unlike with the standard BART model, which has a fixed variance for the leaf node values, 
we adopt the hyperprior $c\sim\mathcal{IG}(a,b)$. 
Following \citet{KleKne2016}, setting $a=1$ mimics a proper but relatively uninformative prior for $1/c$. The hyper-parameter $b$ is set to capture available prior knowledge. Based on the 
observation that $\tilde{z}_i|x_{1:n}\sim \mathcal{N}(0,1+mc)$, approximately $99\%$ of pseudo-observations lie in the interval $\left(-3\sqrt{1+mc},3\sqrt{1+mc}\right)$. We choose $c$ such that $\pm k$-times the standard deviation of the sum of all the leaf node values still lie within this interval, where we set $k=4$. Solving this equation for $c$ gives an optimal value
\begin{equation*}
    k\sqrt{mc} = 3\sqrt{1+mc}\quad \Leftrightarrow\quad c = \frac{9}{(k^2 - 9)m}\,,
\end{equation*}
so that $k = 4$ results in $c^* = \frac{9}{7m}$. The hyper-parameter $b$ is the chosen so that the mode of the prior distribution is $c^*$. The mode of a $\mathcal{IG}(1,b)$ distribution is $\frac{b}{2}$, therefore the optimal $b$ is given by $b^* = \frac{9}{14m}$.

\paragraph{Copula Process Model:}
The copula process model combines the BART copula process above with marginals for the observed response.
Consider  regression data $\{(Y_i,x_i)\}_{i=1}^n$, where $Y_i\in\mathbb{R}$ is the dependent variable. We make the usual regression assumption that $Y_i|x_{1:n} \stackrel{d}{=} Y_i|x_i$ for $1\leq i \leq n$, and denote its distribution function as $F_{Y_i}(y_i|x_i)$ and density as $p_{Y_i}(y_i|x_i)=\frac{d}{dy_i}F_{Y_i}(y_i|x_i)$. 
By Sklar's theorem, the 
joint density of $Y_{1:n}=(Y_1,\ldots,Y_n)^\top$ conditional
on $x_{1:n}$ is 
\begin{equation}\label{eq:sklar}
    p(y_{1:n}|x_{1:n}) = c^\dagger\left(F_{Y_1}(y_1|x_1),\dots,F_{Y_n}(y_n|x_n);x_{1:n}\right)\prod_{i=1}^n p_{Y_i}(y_i|x_i)\,,
\end{equation}
for some $n$-dimensional copula with density $c^\dagger(u_{1:n};x_{1:n})$.

In the copula modeling literature,
a parametric copula is usually selected for $c^\dagger$. 
In this paper, we propose adopting the Gaussian copula process
implicit in the BART model discussed above, and set $c^\dagger(u_{1:n};x_{1:n})=c_{\mbox{\tiny Ga}}(u_{1:n};\Omega(c,\mathcal{E}))$. In the BART model the effect of the covariates is summarized by the tree ensemble $\mathcal{E}$, and it is here too. 
Following~\cite{Klein2019} and~\cite{smithklein21} we also assume
the marginals are invariant to $x_i$, so that $F_{Y_i}(y_i|x_i):=F_Y(y_i)$. This has the advantage that $F_Y$ can be estimated separately from the copula parameters non-parametrically. This assumption also means that it is solely the copula process that captures the impact of the covariates in the distributional predictions. But the assumption does not mean these are homoscedastic, and predictions can have a high degree of distributional heterogeneity over observations as we show later empirically.
Our copula model for~\eqref{eq:sklar} is therefore
\begin{equation}\label{eq:regcop}
    p(y_{1:n}|x_{1:n}) = p( y_{1:n}|c,\mathcal{E}) =
    c_{\mbox{\tiny Ga}}\left(F_{Y}(y_1),\dots,F_{Y}(y_n);\Omega(c,\mathcal{E})\right)
    \prod_{i=1}^np_{Y}(y_i)\,.
\end{equation}
with $p_Y(y)=\frac{d}{dy}F_Y(y)$.

Copulas are also interpretable as transport maps~\citep{kolesarova2008measure}. The following proposition is useful for computing efficiently both the posterior of the copula parameters, and also the predictions from the model.

\begin{proposition}\label{lem:transport}
Let $A$ be a set of realization indices (e.g. $A=\{1:n\}$, $A=\{i\}$ or $A=\{n+1\}$). Consider the element-wise transport map $R$ at these indices given by
\begin{equation}
y_{A}=R(\tilde{z}_{A}) := 
\underline{F_Y}^{-1}\left(\underline{\Phi}(S_{A}\tilde{z}_{A})\right)\,,\label{eq:OT}
\end{equation}
where $S_A=sI_{|A|}$ is the scaling matrix defined above, and $\underline{F_Y}^{-1}$ and $\underline{\Phi}$ denote element-wise application of 
the quantile function $F_Y^{-1}$ and distribution function $\Phi$, respectively. Then the determinant of the Jacobian of the transformation is
\[
\det(J_R):=\left|\frac{dy_{A}}{d\tilde{z}_{A}}\right|=\prod_{i\in A} \frac{s \phi(z_i)}{p_Y(y_i)}\,,
\]
where $z_i=s \tilde z_i$ is the standardized value of the pseudo-response, and $\phi$ is the standard normal density. 
Moreover, if
$\mathds{P}_X$ denotes the probability measure of the distribution of random vector $X$, then
\begin{itemize}
\item[(i)] $R$ pushes  $\mathds{P}_{\widetilde{Z}_{A}|c,\mathcal{E}}$ forward to $\mathds{P}_{Y_{A}|c,\mathcal{E}}$ and is the optimal transport under quadratic loss, and
\item[(ii)] $R$ pushes $\mathds{P}_{\widetilde{Z}_{A}|c,\mathcal{E},\mathcal{M}}$ forward to $\mathds{P}_{Y_{A}|c,\mathcal{E},\mathcal{M}}$\,.
\end{itemize}
\end{proposition}

Part~(i) shows that $R$ is the canonical transformation underlying the Gaussian copula process model with density at~\eqref{eq:regcop}. Part~(ii) is important as we employ it to 
evaluate posterior inference and predictions efficiently using augmentation with the leaf nodes $\mathcal{M}$ below. The proof of Proposition \ref{lem:transport} is provided in the SM, Section A. 

\subsection{Prediction}\label{subsec:pred}
Distributional, point and quantile predictions can all be obtained from the BART copula process model.
 
\paragraph{Distributional Predictions:} The posterior predictive distribution of an unobserved value of the dependent variable $Y_{n+1}$ given observed covariates $x_{n+1}$ can be computed as follows. Let $\eta=\{c,\mathcal{E},\mathcal{M}\}$ denote the copula parameters augmented with the leaf node values, and $A=\{n+1\}$ be a singleton set in~\eqref{eq:OT}. Then it follows from Part~(ii) of Proposition~\ref{lem:transport} and the
regression at~\eqref{eq:regression} (with $\sigma=1$)
that the posterior predictive distribution is
\begin{eqnarray}
 {p}_{n+1}(y_{n+1}|x_{n+1}) &:= &\int p(y_{n+1}|x_{n+1},\eta)p(\eta|y_{1:n})\mbox{d}\eta \nonumber \\
 &= &\int p(\tilde{z}_{n+1}|x_{n+1},\eta)\det(J_R^{-1})
p(\eta|y_{1:n})\mbox{d}\eta \nonumber\\
 &= & \int \phi(\tilde{z}_{n+1}-f_{\mathcal{E},\mathcal{M}}(x_{n+1}))
 \frac{p_Y(y_{n+1})}{s \phi(z_{n+1})}p(\eta|y_{1:n})
\mbox{d}\eta\,,\label{eq:predp}
 \end{eqnarray}
where
\[
\tilde{z}_{n+1}=R^{-1}(y_{n+1})=\Phi^{-1}(F_Y(y_{n+1}))/s \,.
\]

The integral at~\eqref{eq:predp} can be evaluated in the usual Bayesian fashion to obtain an estimator $\widehat{p}_{n+1}(y_{n+1}|x_{n+1})$ using draws from the augmented posterior $\eta^{[k]} \sim p(\eta|y_{1:n})$, for which we suggest an efficient MCMC sampling scheme in Section~\ref{sec:estimation} below. An alternative ``plug in'' estimator replaces $s$ and $f_{\mathcal{E},\mathcal{M}}(x_{n+1})$ with their Monte Carlo estimates 
$\bar{s}=(1+m\bar{c})^{-1/2}$, $\bar{c}=\frac{1}{K} \sum_{k=1}^K c^{[k]}$
and $\bar{f}_{\mathcal{E},\mathcal{M}}(x_{n+1})=\frac{1}{K}\sum_{k=1}^K f_{\mathcal{E}^{[k]},\mathcal{M}^{[k]}}(x_{n+1}) $ constructed from $K$ draws from the augmented posterior.  
Although this under-represents posterior
uncertainty, it is faster to evaluate and we find in our empirical work that it gives very similar distributional predictions.

\paragraph{Mean Predictions:} The posterior predictive mean
\begin{equation}
\dsE(Y_{n+1}|x_{n+1}) := \int \dsE \left(Y_{n+1}|x_{n+1},\eta \right) p(\eta|y_{1:n})d\eta\,,\label{eq:predE}
\end{equation}
(i.e., the mean of~\eqref{eq:predp}) can be used as a point prediction of $Y_{n+1}$ given a new covariate value $x_{n+1}$. The integral can be computed using the draws $\eta^{[k]}$ from the 
augmented posterior to obtain an estimator $\widehat{Y}_{n+1}(x_{n+1})$. However, as before we adopt the faster alternative of  
using the plug-in posterior mean estimators $\bar s$ and $\bar{f}_{\mathcal{E},\mathcal{M}}(x_{n+1})$.
Using Part~(ii) of Proposition~\ref{lem:transport} with 
index set $A=\{n+1\}$, and observing from~\eqref{eq:regression} that $\widetilde{Z}_{n+1} \sim \mathcal{N}(f_{\mathcal{E,M}}(x_{n+1}),1)$, it follows that the term in the integrand of~\eqref{eq:predE} is
\begin{equation*}
\begin{split}
\dsE \left(Y_{n+1}|x_{n+1},\eta \right)&= \int y_{n+1}p(y_{n+1}|x_{n+1},\eta) dy_{n+1}\\
&= \int y_{n+1}
\phi\left(\tilde{z}_{n+1}- f_{\mathcal{E,M}}(x_{n+1})\right)\frac{p_Y(y_{n+1})}{s\phi(z_{n+1})} dy_{n+1}\,.
\end{split}
\end{equation*}
Changing the variable of integration to  $z_{n+1}=s\tilde{z}_{n+1}=\Phi^{-1}(F_Y(y_{n+1}))$ gives
\begin{equation*}
    \dsE \left(Y_{n+1}|x_{n+1},\eta \right)=\int F_Y^{-1}(\Phi(z_{n+1}))
\phi\left(z_{n+1}/s-f_{\mathcal{E,M}}(x_{n+1})\right) dz_{n+1}\,.
\end{equation*}
which is evaluated using univariate numerical integration. Higher order moments  of the predictive density at~\eqref{eq:predp} or other quantities
can be computed in a similar fashion.

\paragraph{Quantile Predictions:}
The posterior predictive quantile ${Q}_{n+1,\alpha}(x_{n+1})$ at quantile level $\alpha\in(0,1)$ of $Y_{n+1}$, given a new observation $x_{n+1}$ is given by 
\begin{eqnarray*}
Q_{n+1,\alpha}(x_{n+1}) &=&\int F_{Y}^{-1}\left(\Phi_1(\varrho_{n+1,\alpha}(x_{n+1}))\right) p(\eta|y_{1:n})d\eta\,, \mbox{ where }\\
	\varrho_{n+1,\alpha}(x_{n+1})&=&\int \left[\Phi^{-1}(\alpha)/s-f_{\mathcal{E,M}}(x_{n+1})\right]  p(\eta|y_{1:n})d\eta\,,
\end{eqnarray*}
and we obtain estimates $\widehat{Q}_{n+1,\alpha}(x_{n+1})$ as we do for point and distributional predictions.
 
\section{Estimation}\label{sec:estimation}
We now discuss how to compute Bayesian inference for the copula process parameters $\eta$, conditional on a first stage
estimate of $p_Y$ given by an adaptive kernel density estimator (KDE). Such two stage estimators are popular 
in the copula literature due to their computational simplicity and strong asymptotic properties~\citep{joe2005}.
Joint estimation to account for posterior uncertainty in $p_Y$ is also possible, but often has minimal impact on inference for $\eta$ and predictive densities, as illustrated in Section~\ref{sec:concreteresults}. 
\subsection{Likelihood}
The likelihood of the copula model is given by~\eqref{eq:regcop}. To evaluate this  requires inversion or factorization of the $(n\times n)$ matrix $\Omega(c,\mathcal{E})=S_{1:n}(cEE^\top + I_n)S_{1:n}$. Direct evaluation involves $O(n^3)$ operations and is infeasible for typical sample sizes $n$.\footnote{The Woodbury identity can be used to factor $\Omega$, but even so it still involves $O(nK^2+K^3)$ computations where $K=\sum_{j=1}^m K^j$ where in our empirical work $m=75$ and $1\leq K^j\leq 5$ are typical, so the computation remains demanding.} To avoid this computation we instead employ an {\em extended likelihood} that augments with the leaf node values $\mathcal{M}$ given by
\begin{eqnarray*}
p(y_{1:n},\mathcal{M}|c,\mathcal{E}) &= &p(y_{1:n}|\eta)\pi(\mathcal{M}|c,\mathcal{E})\\
 &= &p(\tilde{z}_{1:n}|\eta)\prod_{i=1}^n \frac{p_Y(y_i)}{s\phi(z_i)}\prod_{j=1}^m  \phi(M^j;0,cI_{K^j})\,, 
\end{eqnarray*}
where the second line follows from Proposition~\ref{lem:transport} with indices $A=\{1:n\}$.
Recall that the leaf node values were integrated out of the implicit copula model and are not copula parameters, and we stress their re-introduction to aid computation does not change this in any way. From~\eqref{eq:modz},
$p(\tilde{z}_{1:n}|\eta)=\prod_{i=1}^n \phi(\tilde{z}_i-f_{\mathcal{E},\mathcal{M}}(x_i))$, and recall that $\tilde{z}_i=z_i/s=z_i\sqrt{1+mc}$, so that the extended likelihood is 
\begin{eqnarray*}
\lefteqn{p(y_{1:n},\mathcal{M}|c,\mathcal{E}) \propto } \\
&& (1+mc)^{n/2}\prod_{i=1}^n \exp\left\lbrace-\frac{(z_i\cdot \sqrt{1+mc}-f_{\mathcal{E},\mathcal{M}}(x_i))^2-z_i^2}{2}\right\rbrace p_Y(y_i)\prod_{j=1}^m  \phi(M^j;0,cI_{K^j})\,,
\end{eqnarray*}
which is fast to evaluate in $O(n)$ operations.

\subsection{Posterior Evaluation}
The posterior of the copula parameters $\{c,\mathcal{E}\}$ augmented with the leaf node values $\mathcal{M}$ is 
\begin{equation}\label{eq:post}
p(\eta|y_{1:n})\propto p(y_{1:n},\mathcal{M}|c,\mathcal{E})\pi(c,\mathcal{E})\,,
\end{equation}
which is the product of the extended likelihood and prior. Conditional
on the marginal $F_Y$, this can be evaluated efficiently
using MCMC. A key observation is that the conditional posteriors
$p(\mathcal{E}|c,\mathcal{M},y_{1:n})=p(\mathcal{E}|c,\mathcal{M},\tilde{z}_{1:n})$ and  $p(\mathcal{M}|c,\mathcal{E},y_{1:n})=p(\mathcal{M}|c,\mathcal{E},\tilde{z}_{1:n})$, so they are the same as for the Gaussian BART model of the pseudo-response values
specified in Section~\ref{sec:BART}. Therefore, the efficient and scalable steps outlined by~\cite{ChiGeoMcC2010}, adjusted for the leaf node prior at~\eqref{eq:Mprior}, can be used to generate draws from these conditionals. Because these are the key computations in the MCMC scheme, this is why estimation of our copula process is almost as fast as for the original BART model. 

What remains is the generation of the scalar $c$ from its conditional posterior. To do so we 
transform $c$ to the real line by taking the logarithm $\tilde{c} = \log(c)$ 
and use 
Hamiltonian Monte Carlo (HMC) with a leapfrog integrator and dual averaging \citep{Nesterov2007}. This requires evaluation of the logarithm of the conditional posterior and its gradient, for which closed form expressions can be derived as follows. The conditional posterior of $c$ is
\begin{equation*}
\begin{split}
    p(c\ |\ \mathcal{E},\mathcal{M},y_{1:n})&\propto p(y_{1:n},\mathcal{M}|c,\mathcal{E})\pi(c)\\
    &\propto (1+mc)^\frac{n}{2}\cdot \exp\left\lbrace-\frac{\sum_{i=1}^n\left[(z_i\cdot \sqrt{1+mc}-f_{\mathcal{E},\mathcal{M}}(x_i))^2 - z_i^2\right ]}{2}\right\rbrace\\
    &\cdot\left[\prod_{h=1}^m\prod_{k = 1}^{K_h}\frac{1}{\sqrt{c}}\exp\left(-\frac{\mu_{hk}^2}{2c}\right)\right]c^{-a-1}e^{-\frac{b}{c}},    
\end{split}
\end{equation*}
 from which it follows that the log-posterior $l_{\tilde{c}} = \log (p(\tilde{c}\ |\ \mathcal{E},\mathcal{M},y_{1:n}))$ is (up to an additive constant) 
\begin{equation*}
\begin{split}
l_{\tilde{c}} &= \frac{n}{2}\log(1+m\exp(\tilde{c}))-\frac{1}{2}(1+m\exp(\tilde{c}))\sum_{i=1}^nz_i^2+\sqrt{1+m\exp(\tilde{c})}\sum_{i=1}^nz_if_{\mathcal{E},\mathcal{M}}(x_i)\\
&\quad-\frac{K\tilde{c}}{2} - \frac{\mu}{2\exp(\tilde{c})} - a\tilde{c} - \tilde{c} - \frac{b}{\exp(\tilde{c})},
\end{split}
\end{equation*}
where $K = \sum_{h=1}^m K^h$ ($K^h$: Number of leaf nodes in tree $h$) and $\mu = \sum_{h=1}^m\sum_{k=1}^{K^h}\left(M^h_{k}\right)^2$ are the total number of leaf nodes and the sum of the square of all leaf node values, respectively. The gradient of $l_{\tilde{c}}$ is 
\begin{equation*}
\begin{split}
    \frac{\partial l_{\tilde{c}}}{\partial\tilde{c}}&= \frac{nm\exp(\tilde c)}{2(1+m\exp(\tilde c))} -\frac{m\exp(\tilde c)}{2}\sum_{i=1}^nz_i^2 +\frac{m\exp(\tilde c)}{2\sqrt{1+m\exp(\tilde c)}}\sum_{i=1}^nz_if_{\mathcal{E},\mathcal{M}}(x_i)\\
    &-\frac{K}{2}+\frac{\mu}{2\exp(\tilde c)} - a - 1 + \frac{b}{\exp(\tilde c)}. \\
    \end{split}
\end{equation*}

Algorithm~\ref{alg:cpbart} below gives the estimation algorithm for our CP-BART model. 

\begin{algorithm}[th!]
\caption{CP-BART estimation algorithm}
\label{alg:cpbart}
\begin{algorithmic}[1] 

\Require{}{} Data $y_{1:n}$, burn-in size $J_{\text{burn}}$ and Monte Carlo sample size $J_{\text{iter}}$
    \State Fit KDE to the marginal distribution to obtain an estimate $\hat F_Y$ for $F_Y$
    \State Compute the standardized pseudo-responses $z_i = \Phi^{-1}(\hat{F}_Y(y_i))$ for $i=1,\ldots,n$
    \State Initialize $\{\mathcal{E}^{[0]},\mathcal{M}^{[0]},c^{[0]}\}$ to feasible values
    \State Set $s^{[0]} = (1+mc^{[0]})^{-1/2}$ and $\tilde{z}_i^{[0]} = z_i/s^{[0]}$ for $i=1,\ldots,n$
    \For{$j \gets 1$ to $(J_{\text{iter}}+J_{\text{burn}})$}
        \State Draw $\mathcal{E}^{[j]}$ from $p(\mathcal{E}|c^{[j-1]},\mathcal{M}^{[j-1]},\tilde{z}_{1:n}^{[j-1]})$ using the approach in~\cite{ChiGeoMcC2010}
        \State Draw $\mathcal{M}^{[j]}$ from $p(\mathcal{M}|c^{[j-1]},\mathcal{E}^{[j]},\tilde{z}_{1:n}^{[j-1]})$ using the approach in~\cite{ChiGeoMcC2010}
        \State Draw $c^{[j]}$ from $p(c|\mathcal{E}^{[j]},\mathcal{M}^{[j]},\tilde{z}_{1:n}^{[j-1]})$ using HMC
        \State Set $s^{[j]} = (1+mc^{[j]})^{-1/2}$ and $\tilde{z}_i^{[j]} = z_i/s^{[j]}$ for $i=1,\ldots,n$
        \If{$j > J_{\text{burn}}$}
            \State Collect $\lbrace \mathcal{E}^{[j]},\mathcal{M}^{[j]},c^{[j]}\rbrace$
        \EndIf
    \EndFor
   \State Output Monte Carlo Sample from $p(\eta|y_{1:n})$
\end{algorithmic}
\end{algorithm}

\section{Theoretical Results}\label{sec:theory}
In this section, we establish that the fractional posteriors of our CP-BART model for suitable $\rho\in(0,1)$ place their mass in shrinking neighborhoods around the parameters $\vartheta_n^*=(f_n^*,c_n^*)$ of regression function $f_{\mathcal{E},\mathcal{M}}$ and copula parameter $c$ that minimize the average KL divergence between the model-implied conditional densities and the true data generating process. Our Lemma~\ref{lem:klproj} shows that these $\vartheta_n^*$ exist and are unique. The main results are presented in Theorem~\ref{thm1} and Corollary~\ref{cor1}, which establish posterior concentration of regression function and copula parameter, and the corresponding model-implied conditional distributions.

\subsection{Notation and Assumptions}
Let $x_1,\dots,x_n\in\Xc=[0,1]^{p}$ be a fixed design, 
with empirical law $\PX^n=\frac1n\sum_{i=1}^n\delta_{x_i}$, and $p$ fixed. Let $Y_1,\dots,Y_n$ be independent, each $Y_i$ having a conditional distribution  $F_{Y\mid x_i}$ for $i=1,\ldots,n$. 
Let 
$$F_Y^n := \frac1n\sum_{i=1}^n F_{Y|x_i}$$
denote the marginal distribution  of $Y$ under the empirical design measure $\PX^n$ with respective marginal density $p_Y^n$.  
Set $Z_i:=\Phi^{-1}(F^n_Y(Y_i))$ with conditional density $p_0(\cdot\mid x_i)$ which we assume exists and has finite differential-entropy $H(p_0(\cdot\mid x_i))=\mathds{E}_0[-\log p_0(\cdot\mid x_i)]$. Write
\begin{equation*}
      \mu_0(x_i):=\mathds{E}_{0}[Z_i],\qquad \varsigma_0^2(x_i):=\Var_{0}(Z_i),\qquad \gamma_n:=\frac1n\sum_{i=1}^n\varsigma_0^2(x_i).
\end{equation*}
Denote for $\vartheta=(f,c)\in \Theta:= L^{2}(\PX^n)\times (0,\infty)$ the model implied densities of $Z$ and $Y$ by 
\begin{equation*}
  p_{\vartheta}(z\mid x):=s^{-1}\,\phi\bigl(s^{-1}z-f(x)\bigr) \qquad q_{\vartheta}(y\mid x)  = p_{\vartheta}\bigl(\Phi^{-1}(F_{Y}(y))\mid x\bigr)\,  \frac{p_{Y}(y)}{\phi\bigl(\Phi^{-1}(F_{Y}(y))\bigr)}, 
\end{equation*}
where $s=(1+mc)^{-\tfrac{1}{2}}$ as before. 
\begin{remark}
    All of the above quantities,  depend on $n$ through $F^n_Y$. However,  we drop the $n$ from most quantities for readability reasons. We only make the dependence on $n$ of $f^*,s^*,c^*$ and $\gamma$ explicit by writing $f_n^*,s_n^*,c_n^*$ and $\gamma_n$.
\end{remark}
We are interested in finding the minimizer $\vartheta^*_n\in \Theta$ of the average KL divergence between $p_0$ and $p_\vartheta$ given by 
\begin{equation*}
  \mathcal L(\vartheta):=\frac1n\sum_{i=1}^n\mathrm{KL}\bigl(p_0(\cdot\mid x_i)\,\big\|\,p_{\vartheta}(\cdot\mid x_i)\bigr).
\end{equation*}
We make the following assumption on the non-degeneracy of the true $\mu_0$. 
\begin{assumption}\label{ass:M}
$\PX^n(\mu_0\neq0)>0$, i.e.\ $\mu_0(x_i)\neq0$ for at least one $i$.
\end{assumption}
The assumption states that the covariates shift the mean of at least one of the $Z_i$. If this is not the case and $\mu_0\equiv0$, then by Lemma~\ref{lem:klproj} we would have that $c_n^*=0$ which lies outside the support of our prior on $c$.

Assume furthermore, that the splitting probability $\nu$ of the Galton-Watson prior fulfills $\nu\in[1/n,\bar\nu]$ for a fixed $\bar\nu<1$. The CP-BART prior in Section \ref{sec:BART} induces a prior $\Pi$ on $\vartheta$ which we use to define the fractional posterior for $\rho\in(0,1)$ and Borel sets $U\subseteq \Theta$: 
\begin{equation}\label{eq:posttemp}
  \Pi_\rho(U\mid Z_{1:n}):=\frac{\int_U\prod_{i=1}^np_\vartheta(Z_i\mid x_i)^{\rho}\,\Pi(\mathrm d\vartheta)}{\int_\Theta\prod_{i=1}^np_\vartheta(Z_i\mid x_i)^{\rho}\,\Pi(\mathrm d\vartheta)}.
\end{equation}
In the limiting case $\rho\rightarrow1$, \eqref{eq:posttemp} recovers the CP-BART posterior in \eqref{eq:post}. 

We make the following additional assumptions to prove contraction of the KL projection.
\begin{assumption}\label{ass:reg}\
\begin{itemize}
\item[(A2.1)] $\gamma_n\rightarrow\gamma\in(0,1)$.
\item[(A2.2)] $\tau_4:=\sup_{n}\max_{i\le n}\E_{0}W_i^4<\infty$.
\item[(A2.3)] There are $0<\alpha\le1$ and $Q_0\subseteq\{1,\dots,p\}$ with $|Q_0|=q_0\ge1$, both not depending on $n$, such that $f_n^*$ extends from $\{x_1,\dots,x_n\}$ to a function $f_n^*\in\mathcal H^\alpha_p\cap\mathcal C(Q_0)$ on $[0,1]^p$.
The extensions satisfy $B_*:=\sup_n\bigl(\|f_n^*\|_{\mathcal H^\alpha_p}+\|f_n^*\|_\infty\bigr)<\infty$.
For every $n$ the design is $(W,Q_0)$-regular (see Definition B.1 in the SM).
\end{itemize}
\end{assumption}
\begin{assumption}\label{ass:metric}
Let $\|\cdot\|_n$ denote the empirical norm. For  $\rho\in(0,1)$ there is $\kappa_\rho>0$ and $N_0$ such that, for all $n\ge N_0$, all $f:[0,1]^p\to\mathds{R}$ and all $c\in(0,\infty)$:
\begin{equation*}
  -\frac1n\sum_{i=1}^n\log\E_{0}\Bigl[\Bigl(\frac{p_{\vartheta}(Z_i\mid x_i)}{p_{\vartheta^*_n}(Z_i\mid x_i)}\Bigr)^{\rho}\Bigr]
  \ \ge\ \kappa_\rho\min\Bigl\{\|f-f_n^*\|_n^2+(c-c_n^*)^2,\ 1\Bigr\}.
\end{equation*}
\end{assumption}
Assumption~\ref{ass:reg} ensures that the prior mass in neighborhoods around $\vartheta_n^*$ is bounded from below. In particular, (A2.3) are the conditions imposed by \citet{Rokov2020} stated for $f_n^*$.
Assumption~\ref{ass:metric} is a truncated form of the condition introduced by \citet[][Section 3.2]{Bhattacharya2019} and has been assumed before in \citet[][Condition~1]{Syring2023}, with the negative log-likelihood as loss. Without the factor $\kappa_\rho$ it is an averaged form of the strong $\rho$-central condition of \citet[][Definition~7]{Grunwald2020}. 
\subsection{Existence and uniqueness of the KL projection}
\begin{lemma}\label{lem:klproj} Under Assumption \ref{ass:M}, the following holds.
\begin{itemize}
\item[(i)] $\frac1n\sum_{i=1}^n\mu_0(x_i)=0$ and $\frac1n\sum_{i=1}^n\{\mu_0^2(x_i)+\varsigma_0^2(x_i)\}=1$. Hence, $0<\gamma_n=1-\|\mu_0\|_n^2<1$.
\item[(ii)] For every $n>1$, $\vartheta_n^*=(f_n^*,c_n^*)$ with $f_n^*:=\mu_0/\sqrt{\gamma_n}$ and $c_n^*:=(\gamma_n^{-1}-1)/m$ is the unique minimizer of $\mathcal L$ over $\Theta$. This also implies $s^*_n = \sqrt{\gamma_n}$. 
\end{itemize}
\end{lemma}
The proof of Lemma \ref{lem:klproj} is given in the SM, Section C. 

\subsection{Contraction at the KL projection}

 \begin{theorem}\label{thm1}
     Let $\varepsilon_n:=n^{-\alpha/(2\alpha+q_0)}(\log n)^{2\alpha/(2\alpha+q_0)}$ and assume Assumptions~\ref{ass:M}, \ref{ass:reg} and \ref{ass:metric} hold. Then for every sequence $M_n\rightarrow\infty$ and for $n$ large enough we have that
     \begin{equation}\label{eq:thm1}
         \Pi_\rho\Bigl(\vartheta=(f,c):\ \|f-f_n^*\|_n>M_n\varepsilon_n\ \ \text{or}\ \ |c-c_n^*|>M_n\varepsilon_n\ \Big|\ Z_{1:n}\Bigr)\ \le e^{-n\varepsilon_n},
     \end{equation}
 holds on a set with $P^n_0$-probability  $1-\frac{1}{2n\varepsilon_n^2}$, where $P^n_0 = \bigotimes_{i=1}^nP_{Y\ |\ x_i}$.
 \end{theorem}
 Theorem~\ref{thm1} is a consequence of Theorem~3.1 of \citet{Bhattacharya2019} and shows that the fractional posteriors contract around $\vartheta_n^*$ at rate $\varepsilon_n$ for appropriate $\rho\in(0,1)$, which up to the factor $(\log n)^{2\alpha/(2\alpha+q_0)}$ is the minimax rate for $\alpha$-Hölder functions of $q_0$ covariates. The proof of Theorem \ref{thm1} in the SM, Section D verifies the prior-mass condition of \citet{Bhattacharya2019} at $\vartheta_n^*$ and adapts their statement into one about the empirical norm via Assumption\ref{ass:metric}. As a direct consequence, we obtain contraction of the in-sample conditional distributions in average Hellinger distance as formalized in the following Corollary. 
 \begin{corollary}\label{cor1}
     Let $h^2(P,Q):=1-\int\sqrt{\mathrm dP\,\mathrm dQ}$ be the squared Hellinger distance for two distributions $P,Q$. Under the conditions of Theorem~\ref{thm1}, for every $M_n\rightarrow\infty$ and for $n$ large enough we have that
     \begin{equation*}
         \Pi_\rho\Bigl(\vartheta:\ \frac1n\sum_{i=1}^nh^2\bigl(q_\vartheta(\cdot\mid x_i),q_{\vartheta_n^*}(\cdot\mid x_i)\bigr)>M_n^2\varepsilon_n^2\ \Big|\ Z_{1:n}\Bigr)\ \le e^{-n\varepsilon_n},
     \end{equation*}
holds on a set with $P^n_0$-probability $1-\frac{1}{2n\varepsilon_n^2}$.
 \end{corollary}

\begin{remark}
    Throughout this section, we have assumed  the design average marginal distribution of $Y$ to be known. This differs from the practical two-step procedure described in Algorithm~\ref{alg:cpbart}, where this marginal distribution is estimated. The theoretical results therefore allow us to study posterior concentration of the underlying CP-BART structure induced by the Gaussian copula process in isolation. 
\end{remark}

\section{Simulation Study}\label{sec:simulations}
We undertake a simulation study to demonstrate the efficacy of using CP-BART for distributional prediction.

\subsection{Simulation Design}
The data generating processes (DGPs) use the five-dimensional Friedman function~\citep{Friedman1983}
\begin{equation*}
    f({x}) := 10\sin{(\pi x_1x_2)}+20\left(x_3 - \frac{1}{2}\right)^2 + 10x_4 + 5x_5,
\end{equation*} which exhibits nonlinearity and variable interactions. Let $x=(x_1,\ldots,x_5)^\top$. The covariates values are first drawn $x_i\sim \mathcal{N}(0,\Sigma)$ independently for $i=1,\ldots,n$, with a Toeplitz covariance structure $\Sigma_{lj} = \rho^{|l-j|}$, and then standardized to lie in the unit interval. Setting $\rho=0.3$ produces a mildly correlated covariate vector. 
We further normalize the function by approximations of its mean and twice its standard deviation
\begin{equation*}
     f^\ast(x) := \frac{f(x)-15}{2\cdot\sqrt{5.5}}.
\end{equation*}
We sample from the following three distributions:
\begin{equation*}
\begin{split}
    \text{Case 1, Normal:}\hspace{50pt} &Y = f^\ast(x) + \varepsilon, \hspace{10pt} \varepsilon\sim\mathcal{N}(0,r_1^2).\\
    \text{Case 2, Implicit Copula:}\hspace{10pt} &Y = F^{-1}_{\text{Gamma}}(\Phi(z); 3,2),\hspace{10pt} z = f^\ast(x)+\varepsilon, \hspace{10pt} \varepsilon\sim\mathcal{N}(0,r_2^2).\\
    \text{Case 3, Gamma:}\hspace{48pt} &Y = F^{-1}_{\text{Gamma}}\left(u;\frac{f(x)}{\sqrt{5.5}},r_3\right),\hspace{10pt} u\sim U(0,1).
\end{split}
\end{equation*}
The scale values $r_l^2$ for $l=1,2,3$ are chosen so that the signal-to-noise ratio (SNR) is approximately four in each case, making it simpler to compare results. The SNR is defined here as
\begin{equation*}
    \text{SNR} = \frac{\text{range}(\mathbb{E}(Y\ |\ x))}{\left(\int\text{Var}(Y\ |\ x)dx\right)^\frac{1}{2}},
\end{equation*}
from which the scales are computed to be $r_1^2=2.0787$, $r_2^2=1.6$ and $r_3^2=0.2601$; see the SM, Section F for details.

We simulate $R=100$ replicates of training data from all three cases with sample sizes $n = 250,500,1000$. To measure the accuracy, for each case we generate one further test dataset $\{(y_i,x_i);i\in D\}$ with $|D|=n_\text{test}=250$ observations. Results are based on $m=75$ trees and 5000 MCMC iterations, including a burn-in phase of 1000 iterations giving us 4000 samples in total.

\subsection{Evaluation Metrics}
For each test dataset 
 we compute the following three metrics. 
First, to assess the accuracy of point predictions $\widehat{Y}_i(x_i)$ of the dependent variable, we compute the root mean square error,
\begin{equation*}
\text{RMSE}= \sqrt{\frac{1}{n_\text{test}}\sum_{i\in D }(\widehat{Y}_i(x_i) -y_i)^2}\,.
\end{equation*} 
Second, to assess the accuracy of the predictive density $\widehat{p}_i(y_i|x_i)$ for the conditional distribution $Y_i|x_i$, we compute the mean log-score,
\begin{equation*}
   \text{LS} = -\frac{1}{n_\text{test}}\sum_{i\in D}\log\left\lbrace \widehat{p}_{i}(y_i|x_i)\right\rbrace,
\end{equation*} 
orientated so that lower values indicate greater accuracy. 
Third, because the true quantiles $Q_{i,\alpha}(x_i)$ can be computed from the DGP's by simulation, we compute the RMSE between these and the predicted quantiles as a summary of their accuracy. That is, for a given quantile $\alpha$ we compute
\begin{equation*}
\text{QRMSE}_\alpha= \sqrt{\frac{1}{n_\text{test}}\sum_{i\in D }\left(\widehat{Q}_{i,\alpha}(x_i) -Q_{i,\alpha}(x_i)\right)^2}\,.
\end{equation*} 

Fourth, following \cite{LiLinMur2023}, for estimation using MCMC we compute coverage of the 95\% posterior predictive probability intervals for the quantiles $Q_{i,\alpha}(x_i)$ as follows.
For a given observation $i \in D$ and value
$\alpha$, we use a sample from the parameter posterior to compute a sample $\{Q_{i,\alpha}(x_i)^{[1]},\ldots,Q_{i,\alpha}(x_i)^{[J_{iter}]}\}$ from the posterior of the quantiles. From the ordered sample, 95\% probability intervals can be evaluated.
Finally, we report the coverage of these intervals across the test data at $\alpha=0.25$, 0.5 and 0.75.

\subsection{Benchmarks}
The benchmarks for the proposed CP-BART model include the (i)~original BART model with Gaussian errors~of~\cite{ChiGeoMcC2010}, (ii)~SBART-DS model of \citet{LiLinMur2023}, and (iii)~``random forest conditional distribution estimation'' (RFCDE) approach of \citet{rfcde}. 
Mean, density and quantile predictions for CP-BART are obtained as described in Section~\ref{sec:BARTCPM}. 
Predictions for BART are obtained in a similar manner using draws from the standard MCMC scheme. 
The SBART-DS model estimates conditional densities by tilting a base model (e.g., Gaussian) using a flexible function modeled with soft Bayesian additive regression trees. All three models are estimated
using MCMC, from which coverage of the predicted quantiles can be computed. 

RFCDE uses a random forest model designed for conditional density estimation. It 
modifies the traditional random forest algorithm by optimizing tree splits using a loss function tailored to conditional density estimation, from which mean and quantile predictions can also be evaluated. However, this approach does not quantify the epistemic uncertainty 
of the estimates (e.g. by using either a posterior or sampling distribution) so that coverage of the predictions cannot be evaluated.
For both SBART-DS and RFCDE we use the code provided by the authors.

\begin{table}[tbh]
\begin{center}
\caption{Predictive Accuracy Metrics for Simulation Case 1.}
\label{tab:case1}
{\footnotesize
\setlength{\tabcolsep}{6pt}
\begin{tabular}{
l
S[table-format=1.2]
S[table-format=1.2]
S[table-format=1.2]
S[table-format=1.2]
S[table-format=1.2]
S[table-format=1.2]
S[table-format=1.2]
S[table-format=1.2]
}
\toprule
\multirow{2}{*}{Method} 
& \multicolumn{3}{c}{QRMSE} 
& \multicolumn{3}{c}{Quantile Coverage} 
& \multicolumn{1}{c}{RMSE} 
& \multicolumn{1}{c}{LS} \\
\cmidrule(lr){2-4} \cmidrule(lr){5-7}
& {$\alpha=0.25$} & {$\alpha=0.50$} & {$\alpha=0.75$}
& {$\alpha=0.25$} & {$\alpha=0.50$} & {$\alpha=0.75$}
&  &  \\
\midrule
\addlinespace[0.3em]
\multicolumn{9}{@{}l}{\textbf{n = 250}} \\[0.25em]
\hspace{1em}BART & \textbf{0.65} & \textbf{0.65} & \textbf{0.65} & 0.97 & 0.97 & 0.97 & \textbf{0.65} & 2.19\\
\hspace{1em}CP-BART & 0.66 & 0.66 & 0.66 & \textbf{0.97} & \textbf{0.97} & \textbf{0.96} & 0.66 & 2.21\\
\hspace{1em}RFCDE & 0.97$^{***}$ & 0.95$^{***}$ & 0.96$^{***}$ & {--} & {--} & {--} & 0.82$^{***}$ & 2.34$^{***}$\\
\hspace{1em}SBART-DS & 0.71$^{***}$ & 0.71$^{***}$ & 0.72$^{***}$ & 0.83 & 0.82 & 0.81 & 0.71$^{***}$ & \textbf{2.18}\\ 
\addlinespace[0.3em]
\multicolumn{9}{@{}l}{\textbf{n = 500}} \\[0.25em]
\hspace{1em}BART & \textbf{0.59} & \textbf{0.58} & \textbf{0.58} & 0.97 & 0.97 & 0.97 & 0.58 & 2.18\\
\hspace{1em}CP-BART & 0.59 & 0.59 & 0.59 & \textbf{0.96} & \textbf{0.96} & \textbf{0.96} & \textbf{0.58} & 2.19\\
\hspace{1em}RFCDE & 0.89$^{***}$ & 0.86$^{***}$ & 0.88$^{***}$ & {--} & {--} & {--} & 0.76$^{***}$ & 2.32$^{***}$\\
\hspace{1em}SBART-DS & 0.66$^{***}$ & 0.67$^{***}$ & 0.67$^{***}$ & 0.75 & 0.74 & 0.73 & 0.67$^{***}$ & \textbf{2.17}\\ 
\addlinespace[0.3em]
\multicolumn{9}{@{}l}{\textbf{n = 1000}} \\[0.25em]
\hspace{1em}BART & 0.53$^{*}$ & 0.53 & 0.53 & 0.98 & 0.98 & 0.98 & 0.53$^{*}$ & \textbf{2.17}\\
\hspace{1em}CP-BART & \textbf{0.52} & \textbf{0.52} & \textbf{0.52} & \textbf{0.97} & \textbf{0.97} & \textbf{0.97} & \textbf{0.52} & 2.18\\
\hspace{1em}RFCDE & 0.82$^{***}$ & 0.80$^{***}$ & 0.82$^{***}$ & {--} & {--} & {--} & 0.70$^{***}$ & 2.30$^{***}$\\
\hspace{1em}SBART-DS & 0.64$^{***}$ & 0.63$^{***}$ & 0.64$^{***}$ & 0.71 & 0.70 & 0.69 & 0.64$^{***}$ & 2.17\\
\addlinespace[0.3em]
\multicolumn{9}{@{}l}{\textbf{n = 5000}} \\[0.25em]
\hspace{1em}BART & 0.44 & 0.44 & 0.44 & 0.97 & 0.97 & 0.97 & 0.44 & \textbf{2.15}\\
\hspace{1em}CP-BART & \textbf{0.44} & \textbf{0.44} & \textbf{0.44} & \textbf{0.95} & \textbf{0.95} & \textbf{0.94} & \textbf{0.44} & 2.16\\
\hspace{1em}RFCDE & 0.73$^{***}$ & 0.71$^{***}$ & 0.73$^{***}$ & {--} & {--} & {--} & 0.63$^{***}$ & 2.27$^{***}$\\
\bottomrule
\end{tabular}
}
\end{center}
Note: Methods and metrics are outlined in the text. Mean QRMSE, RMSE and LS values across all replicates are reported, with low values indicating higher accuracy and the lowest values in bold. Mean metrics significantly greater/smaller than CP-BART are marked with ``{\small ***/+++}'' (0.1\% level), ``{\small **/++}'' (1\% level) and ``{\small */+}'' (5\% level). Quantile coverage is at the 95\% level for three quantile values $\alpha=0.25, 0.5$ and $0.75$. Coverage is not reported for RFCDE as the method does not quantify epistemic uncertainty, while SBART-DS is not reported for the sample size $n=5000$ due to high computational demand.   
\end{table}

\subsection{Results}
Tables~\ref{tab:case1}, \ref{tab:case2}, and~\ref{tab:case3} present the evaluation metrics for the three cases, with results given for different sample sizes and methods. The ranking of the methods by speed (from fastest to slowest) is RFCDE, BART, CP-BART and SBART-DS. For example, predictions were obtained for a dataset of size $n=5000$ for the methods in this order in an average of 14.5s, 41.8s, 47.5s and 2092s; see Table G.4 in the SM for a comparison of speed with increasing $n/p$. The fast speed of CP-BART is because Algorithm~\ref{alg:cpbart} involves only inexpensive additional computations over-and-above that for BART, which is well-known for its scalability. 
Because of its higher computational time, results are not reported for SBART-DS when $n=5000$. 

\begin{table}[tbh]
\begin{center}
\caption{Predictive Accuracy Metrics for Simulation Case 2.}
\label{tab:case2}
\footnotesize
\setlength{\tabcolsep}{6pt}
\begin{tabular}{
l
S[table-format=1.2]
S[table-format=1.2]
S[table-format=1.2]
S[table-format=1.2]
S[table-format=1.2]
S[table-format=1.2]
S[table-format=1.2]
S[table-format=1.2]
}
\toprule
\multirow{2}{*}{Method} 
& \multicolumn{3}{c}{$\text{RMSE}_Q$} 
& \multicolumn{3}{c}{Quantile Coverage} 
& \multicolumn{1}{c}{RMSE} 
& \multicolumn{1}{c}{LS} \\
\cmidrule(lr){2-4} \cmidrule(lr){5-7}
& {$\alpha=0.25$} & {$\alpha=0.50$} & {$\alpha=0.75$}
& {$\alpha=0.25$} & {$\alpha=0.50$} & {$\alpha=0.75$}
&  &  \\
\midrule
\addlinespace[0.3em]
\multicolumn{9}{@{}l}{\textbf{n = 250}} \\[0.25em]
\hspace{1em}BART & 0.39$^{***}$ & 0.57$^{***}$ & 0.66$^{***}$ & 0.99 & 0.94 & 0.89 & 0.40$^{***}$ & 1.95$^{***}$\\
\hspace{1em}CP-BART & \textbf{0.21} & \textbf{0.33} & \textbf{0.48} & \textbf{0.96} & \textbf{0.96} & \textbf{0.96} & \textbf{0.35} & \textbf{1.56}\\
\hspace{1em}RFCDE & 0.33$^{***}$ & 0.52$^{***}$ & 0.76$^{***}$ & {--} & {--} & {--} & 0.49$^{***}$ & 1.76$^{***}$\\
\hspace{1em}SBART-DS & 0.25$^{***}$ & 0.42$^{***}$ & 0.58$^{***}$ & 0.76 & 0.71 & 0.81 & 0.40$^{***}$ & 1.61$^{***}$\\ 
\addlinespace[0.3em]
\multicolumn{9}{@{}l}{\textbf{n = 500}} \\[0.25em]
\hspace{1em}BART & 0.39$^{***}$ & 0.55$^{***}$ & 0.61$^{***}$ & 0.98 & 0.91 & 0.87 & 0.36$^{***}$ & 1.95$^{***}$\\
\hspace{1em}CP-BART & \textbf{0.18} & \textbf{0.29} & \textbf{0.43} & \textbf{0.95} & \textbf{0.95} & \textbf{0.95} & \textbf{0.31} & \textbf{1.55}\\
\hspace{1em}RFCDE & 0.31$^{***}$ & 0.48$^{***}$ & 0.74$^{***}$ & {--} & {--} & {--} & 0.47$^{***}$ & 1.73$^{***}$\\
\hspace{1em}SBART-DS & 0.23$^{***}$ & 0.37$^{***}$ & 0.50$^{***}$ & 0.69 & 0.71 & 0.84 & 0.35$^{***}$ & 1.59$^{***}$\\ 
\addlinespace[0.3em]
\multicolumn{9}{@{}l}{\textbf{n = 1000}} \\[0.25em]
\hspace{1em}BART & 0.39$^{***}$ & 0.54$^{***}$ & 0.59$^{***}$ & 0.96 & 0.88 & 0.82 & 0.33$^{***}$ & 1.94$^{***}$\\
\hspace{1em}CP-BART & \textbf{0.16} & \textbf{0.25} & \textbf{0.36} & \textbf{0.96} & \textbf{0.96} & \textbf{0.96} & \textbf{0.27} & \textbf{1.54}\\
\hspace{1em}RFCDE & 0.29$^{***}$ & 0.46$^{***}$ & 0.70$^{***}$ & {--} & {--} & {--} & 0.45$^{***}$ & 1.70$^{***}$\\
\hspace{1em}SBART-DS & 0.20$^{***}$ & 0.33$^{***}$ & 0.46$^{***}$ & 0.69 & 0.74 & 0.85 & 0.33$^{***}$ & 1.56$^{***}$\\ 
\addlinespace[0.3em]
\multicolumn{9}{@{}l}{\textbf{n = 5000}} \\[0.25em]
\hspace{1em}BART & 0.36$^{***}$ & 0.49$^{***}$ & 0.54$^{***}$ & 0.92 & 0.71 & 0.65 & 0.28$^{***}$ & 1.93$^{***}$\\
\hspace{1em}CP-BART & \textbf{0.13} & \textbf{0.21} & \textbf{0.31} & \textbf{0.95} & \textbf{0.95} & \textbf{0.95} & \textbf{0.22} & \textbf{1.54}\\
\hspace{1em}RFCDE & 0.24$^{***}$ & 0.42$^{***}$ & 0.68$^{***}$ & {--} & {--} & {--} & 0.44$^{***}$ & 1.68$^{***}$\\
\bottomrule
\end{tabular}
\end{center}
See note to Table~\ref{tab:case1} for table description.
\end{table}

\begin{table}[tbh]
\begin{center}
\caption{Predictive Accuracy Metrics for Simulation Case 3.}
\label{tab:case3}
\footnotesize
\setlength{\tabcolsep}{6pt}
\begin{tabular}{
l
S[table-format=1.2]
S[table-format=1.2]
S[table-format=1.2]
S[table-format=1.2]
S[table-format=1.2]
S[table-format=1.2]
S[table-format=1.2]
S[table-format=1.2]
}
\toprule
\multirow{2}{*}{Method} 
& \multicolumn{3}{c}{$\text{RMSE}_Q$} 
& \multicolumn{3}{c}{Quantile Coverage} 
& \multicolumn{1}{c}{RMSE} 
& \multicolumn{1}{c}{LS} \\
\cmidrule(lr){2-4} \cmidrule(lr){5-7}
& {$\alpha=0.25$} & {$\alpha=0.50$} & {$\alpha=0.75$}
& {$\alpha=0.25$} & {$\alpha=0.50$} & {$\alpha=0.75$}
&  &  \\
\midrule
\addlinespace[0.3em]
\multicolumn{9}{@{}l}{\textbf{n = 250}} \\[0.25em]
\hspace{1em}BART & 1.17$^{***}$ & 1.21$^{***}$ & 1.31$^{***}$ & 0.99 & 0.98 & 0.98 & 1.10$^{***}$ & 2.67$^{***}$\\
\hspace{1em}BART-cu & 0.82$^{***}$ & 0.98$^{***}$ & 1.19$^{***}$ & 0.99 & 0.97 & 0.93 & 0.85$^{**}$ & 2.63$^{***}$\\
\hspace{1em}CP-BART & \textbf{0.53} & \textbf{0.74} & \textbf{1.04} & 0.97 & \textbf{0.96} & \textbf{0.94} & \textbf{0.80} & \textbf{2.44}\\
\hspace{1em}RFCDE & 0.89$^{***}$ & 1.11$^{***}$ & 1.49$^{***}$ & {--} & {--} & {--} & 1.00$^{***}$ & 2.60$^{***}$\\
\hspace{1em}SBART-DS & 0.64$^{***}$ & 0.87$^{***}$ & 1.16$^{***}$ & \textbf{0.94} & 0.90 & 0.86 & 0.88$^{***}$ & 2.48$^{***}$\\ 
\addlinespace[0.3em]
\multicolumn{9}{@{}l}{\textbf{n = 500}} \\[0.25em]
\hspace{1em}BART & 1.11$^{***}$ & 1.13$^{***}$ & 1.20$^{***}$ & 0.98 & 0.98 & 0.98 & 1.00$^{***}$ & 2.66$^{***}$\\
\hspace{1em}BART-cu & 0.76$^{***}$ & 0.89$^{***}$ & 1.07$^{***}$ & 0.98 & \textbf{0.96} & 0.93 & 0.74$^{***}$ & 2.62$^{***}$\\
\hspace{1em}CP-BART & \textbf{0.45} & \textbf{0.63} & \textbf{0.91} & \textbf{0.97} & 0.96 & \textbf{0.94} & \textbf{0.69} & \textbf{2.43}\\
\hspace{1em}RFCDE & 0.84$^{***}$ & 1.05$^{***}$ & 1.39$^{***}$ & {--} & {--} & {--} & 0.93$^{***}$ & 2.59$^{***}$\\
\hspace{1em}SBART-DS & 0.60$^{***}$ & 0.81$^{***}$ & 1.05$^{***}$ & 0.91 & 0.88 & 0.86 & 0.81$^{***}$ & 2.47$^{***}$\\ 
\addlinespace[0.3em]
\multicolumn{9}{@{}l}{\textbf{n = 1000}} \\[0.25em]
\hspace{1em}BART & 1.04$^{***}$ & 1.07$^{***}$ & 1.14$^{***}$ & 0.97 & 0.97 & \textbf{0.96} & 0.92$^{***}$ & 2.65$^{***}$\\
\hspace{1em}BART-cu & 0.72$^{***}$ & 0.85$^{***}$ & 1.03$^{***}$ & \textbf{0.96} & 0.94 & 0.88 & 0.66$^{**}$ & 2.62$^{***}$\\
\hspace{1em}CP-BART & \textbf{0.41} & \textbf{0.58} & \textbf{0.83} & 0.97 & \textbf{0.95} & 0.93 & \textbf{0.64} & \textbf{2.43}\\
\hspace{1em}RFCDE & 0.77$^{***}$ & 0.98$^{***}$ & 1.30$^{***}$ & {--} & {--} & {--} & 0.87$^{***}$ & 2.57$^{***}$\\
\hspace{1em}SBART-DS & 0.56$^{***}$ & 0.76$^{***}$ & 1.01$^{***}$ & 0.88 & 0.86 & 0.85 & 0.77$^{***}$ & 2.46$^{***}$\\ 
\addlinespace[0.3em]
\multicolumn{9}{@{}l}{\textbf{n = 5000}} \\[0.25em]
\hspace{1em}BART & 0.83$^{***}$ & 0.88$^{***}$ & 0.97$^{***}$ & 0.93 & 0.90 & 0.86 & 0.69$^{***}$ & 2.63$^{***}$\\
\hspace{1em}BART-cu & 0.62$^{***}$ & 0.74$^{***}$ & 0.91$^{***}$ & 0.89 & 0.83 & 0.74 & 0.52 & 2.61$^{***}$\\
\hspace{1em}CP-BART & \textbf{0.32} & \textbf{0.46} & \textbf{0.69} & \textbf{0.96} & \textbf{0.95} & \textbf{0.91} & \textbf{0.52} & \textbf{2.43}\\
\hspace{1em}RFCDE & 0.70$^{***}$ & 0.88$^{***}$ & 1.15$^{***}$ & {--} & {--} & {--} & 0.79$^{***}$ & 2.57$^{***}$\\
\bottomrule
\end{tabular}
\end{center}
See note to Table~\ref{tab:case1} for table description.  
\end{table}

In Case~1 the DGP has an additive Gaussian error, matching the BART model, so that this method is most accurate in Table~\ref{tab:case1}. However, of the three distributional regression methods, CP-BART is most accurate by all measures, except for LS where SBART-DS is superior.
In Case~2 the data are generated from a transformation of a latent nonlinear regression with Gaussian disturbances. This is a Gaussian copula process model, so that CP-BART should perform well here. It does so in Table~\ref{tab:case2}, dominating the other approaches on all metrics and at all sample sizes. In Case~3 the DGP has a highly non-Gaussian response distribution that varies with covariate values, but is not a copula process model. As expected, in Table~\ref{tab:case3} the three distributional regression methods clearly out-perform the BART model which only allows the conditional mean to vary with covariates. Of these, CP-BART dominates throughout, followed by SBART-DS and then RFCDE.

We make two additional observations on CP-BART. First, it estimates the conditional quantiles very well, both in terms of point forecast accuracy measured by QRMSE and with well-calibrated coverage. This is despite the method not being a tailored quantile regression approach. 
Second, similar to the original BART model, we tried fixing the scale of the leaf nodes $c$ to a constant ($\frac{9}{7m}$, see Section \ref{sec:BARTCPM} for details), but this reduced the predictive accuracy. This is because $c$ is crucial when estimating the copula process parameter matrix $\Omega(c,\mathcal{E})$, and its posterior can differ meaningfully for any given dataset. Likewise we tried sampling the scale of the leaf nodes explicitly in BART using HMC however this again seems to reduce predictive accuracy. The corresponding evaluation metrics can be found in Tables G.1--G.3 of the SM.
\FloatBarrier

\section{Real World Data Illustrations}\label{sec:real}
We evaluate our CP-BART and the benchmark methods considered in Section \ref{sec:simulations} on the five real world regression datasets listed in Table~\ref{tab:UCIdatasets}. All datasets but the Rain data (which is from the \textsf{R}-package \texttt{disttree}) were obtained from the UCI repository. The datasets vary in sample size and number of covariates, so that they represent a variety of circumstances. All are where the relationship between the covariates and the response is possibly nonlinear and not additive, with a response distribution that is non-Gaussian and may be covariate dependent. Thus, distributional regression tree models are an attractive choice. Further implementation details are given in SM, Section H. 

\begin{table}[thb]
\label{tab:UCIdatasets}
\begin{center}
\caption{Five Real World Regression Datasets}
\small
\begin{tabular}{lllll}
\toprule
Dataset & Source & $n$ & $p$ & Response Variable ($y$) \\
\midrule
Housing & UCI-NoID & 506  & 12 & Median value of owner-occupied homes (USD 1,000)\\
Rain & R-disttree  & 867  & 82  & Power transformed observed total precipitation ($\text{mm}^\frac{1}{1.6}$)\\
Concrete & UCI-ID165 & 1,030 & 8 & Concrete compressive strength (MPa)\\
Bike  & UCI-ID275 & 17,379 & 12 & Bike rentals (count)\\
Music & UCI-ID203 & 515,345 & 90  & Release year of songs\\
\bottomrule
\end{tabular}
\end{center}
\end{table}

\subsection{CP-BART Results}
Each dataset was fit using CP-BART and we make three observations. First, $c$ is a 
key parameter in the copula process, and  plots of MCMC draws of it are an effective diagnostic of convergence of the sampler. These are given in Figure H.1 in the SM, and visual inspection suggests convergence is fast for the three smaller datasets, although more draws are necessary for the large Bike and Music datasets. Second, Figure~\ref{fig:rainbikedensity} plots the predictive densities for three representative observations in the Rain and Bike datasets. This includes the posterior mean of the predictive densities, and there is strong variability in the shape of the predictive densities across observations. Thus, accounting for the covariate values through the copula process, and not through the marginal $p_Y$, at~\eqref{eq:regcop} is impactful. Third, 
also plotted are 90\% posterior probability intervals for the densities. This is obtained by evaluating the predictive densities at draws from $p(\eta|y)$, instead of integrating out posterior uncertainty as in~\eqref{eq:predp}. The densities for the Rain dataset exhibit high epistemic uncertainty, which reflects its low sample size to variable ratio. In contrast, the posterior intervals are narrow for the Bike dataset, reflecting the high sample size to variable ratio for this case.

\begin{figure}[htb]
    \centering
    \includegraphics[width=1\linewidth]{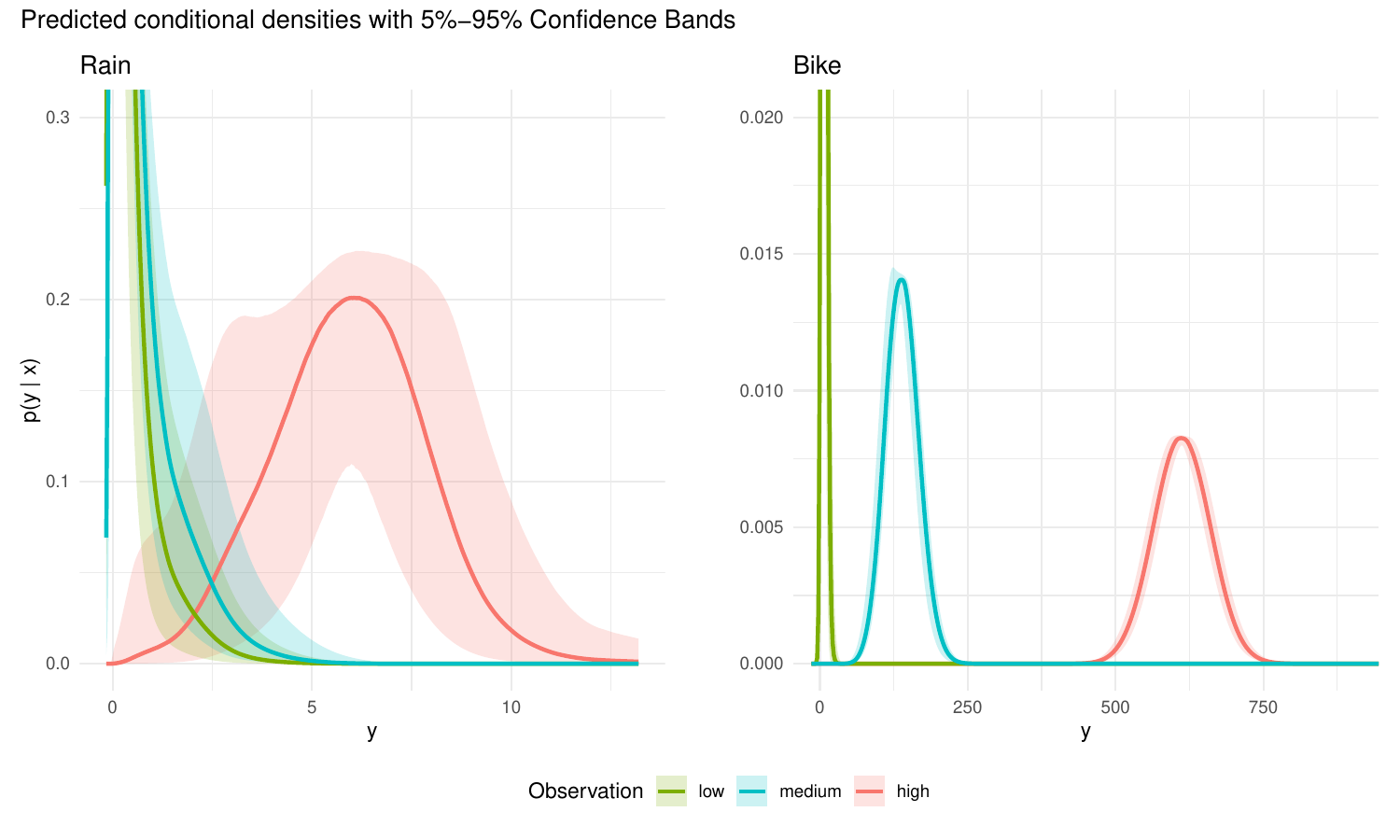}
    \caption{Predictive densities from CP-BART for the Rain (lefthand panel) and Bike (righthand panel) datasets. For each dataset, predictions are made for the three observations corresponding the 5th, 50th and 95th percentiles of the response variable. Solid lines depict the posterior median of the predictive densities, while the shaded areas give the 90\% posterior predictive probability intervals.}
    \label{fig:rainbikedensity}
\end{figure}

\subsection{Comparison of Predictive Accuracy}
To compare the predictive accuracy of the four models we compute the predictive (i)~pinball loss (also called the quantile score), (ii)~continuous ranked probability score~\citep[CRPS;][]{GneitingRaftery2007} and (iii)~log-scores, using 10-fold cross-validation as follows. For a given dataset, the data are randomly partitioned into ten (approximately) equal sized disjoint sub-samples. 
Each sub-sample has observation indices $B^{(k)}$ for $k=1,\ldots,10$, so that $\bigcup_{k=1}^{10}B^{(k)}=\{1,2,\ldots,n\}$. For each sub-sample $k$, a model is fit using the other nine sub-samples as training data, and the predictive densities $\hat{p}_i(y_i|x_i)$ and quantile functions $\widehat{Q}_{i,\alpha}(x_i)$ evaluated for $i\in B^{(k)}$ which are treated as test data. This is repeated for each partition $k=1,\ldots,10$ and the metrics evaluated across all observations.

The pinball loss is defined at quantile $\alpha$ as $\text{PL}(\alpha)=\frac{1}{n}\sum_{i=1}^n \text{PL}_{i,\alpha}(y_i|x_i)$, where
\begin{equation*}
    \text{PL}_{i,\alpha}(y_i|x_i) := \left(\mathds{1}\left(y_i<\widehat{Q}_{i,\alpha}(x_i)\right)-\alpha\right)\cdot\left(\widehat{Q}_{i,\alpha}(x_i) - y_i\right)\,,
\end{equation*}
and the indicator function $\mathds{1}(X)=1$ if $X$ is true, and zero otherwise. The CRPS is the integral of the pinball loss over all quantiles
\[
\text{CRPS}(y_i)=2\int_{0}^1 \text{PL}_{i,\alpha}(y_i|x_i)d\alpha\,, 
\]
and we compute the mean over all observations. Cross-validation is used for four of the datasets, but not for the largest Music dataset because it has a pre-specified train/test split (463,715/51,630) that avoids the ``producer effect'' meaning no songs are from the same artist in both the test and training data.

\begin{table}[tbh]
\begin{center}
\caption{Predictive Accuracy for the Real World Datasets}
\label{tab:realDataresults}
{\small \begin{threeparttable}
\begin{tabular}{llllll}
\toprule
\textbf{Method} & \textbf{Housing} & \textbf{Rain} & \textbf{Concrete} & \textbf{Bike} & \textbf{Music} \\
\midrule
\addlinespace[0.25em]
\multicolumn{6}{l}{\textbf{CRPS}} \\
\addlinespace[0.25em]
\hspace{1em}BART      & 1.53              & 0.85$^{***}$       & 2.17$^{*}$             & 18.95$^{***}$       & 4.88$^{***}$ \\
\hspace{1em}CP-BART   & \textbf{1.49}     & \textbf{0.68}      & \textbf{1.99}     & \textbf{17.31}      & \textbf{4.41} \\
\hspace{1em}RFCDE     & 1.80$^{**}$        & 0.70               & 3.22$^{***}$      & 19.70$^{***}$       & 4.65$^{***}$ \\
\hspace{1em}SBART-DS  & 2.00$^{***}$      & \text{NI}            & 4.87$^{***}$      &\text{CD} &\text{CD} \\
\addlinespace[0.4em]
\multicolumn{6}{l}{\textbf{Log-score}} \\
\addlinespace[0.25em]
\hspace{1em}BART      & 2.73              & 1.84$^{***}$       & 2.92              & 5.16$^{***}$        & 3.62$^{***}$ \\
\hspace{1em}CP-BART   & 2.54              & \textbf{0.83}      & \textbf{2.84}     & \textbf{4.67}       & \textbf{3.24} \\
\hspace{1em}RFCDE     & \textbf{2.49}     & \text{NI}      & 3.18$^{***}$        & \text{NI}                & 3.25$^{*}$   \\
\hspace{1em}SBART-DS  & 2.67              &\text{NI}                & 3.55$^{***}$      &\text{CD}                  &\text{CD}\\
\bottomrule
\end{tabular}
\end{threeparttable}
}
\end{center}
{\small Note: Mean metric values are reported, computed using 10-fold CV as described in the text. Lower values correspond to higher accuracy, with the lowest in bold. Mean metrics significantly greater/smaller than CP-BART are marked with ``{\small ***/+++}'' (0.1\% level), ``{\small **/++}'' (1\% level) and ``{\small */+}'' (5\% level). The acronym ``CD'' indicates computationally difficult to evaluate on standard machines, while ``NI'' indicates numerical instabilities were encountered when employing the code.
}
\end{table}

Table~\ref{tab:realDataresults} reports the predictive CRPS and LS values. Results are not given for SBART-DS for the two largest datasets (Bike and Music) due to computational hurdles, and for the Rain dataset where the the code encountered numerical instabilities due to the challenging $n$-to-$p$ ratio. CP-BART is most accurate throughout when measured by CRPS, 
and for three datasets by LS. Only for two datasets (Housing and Music)  RFCDE is more accurate than CP-BART, and with Housing the scores are not significantly different.

Figure~\ref{fig:pinball} plots the pinball loss functions $\text{PL}(\alpha)$. These show that the CP-BART provides more accurate predictions than the three benchmark methods at all quantiles for all five datasets. To illustrate why this is case, Figure~\ref{fig:housepred} plots the predictive densities for each method and four observations in the Housing data. The observations correspond to one of the cheapest, one of the most expensive and a median priced property tract, along with that with the lowest crime rate. Care was taken to fit the models to the data excluding these observations, so that the predictions are out-of-sample. All four densities from CP-BART are smoother, better located, and/or sharper than the benchmarks. Their shape also varies greatly over observation, showing how the copula process is responsive to covariate values.

\begin{figure}[thbp]
    \centering
    \includegraphics[angle=-90,width=0.9\linewidth]{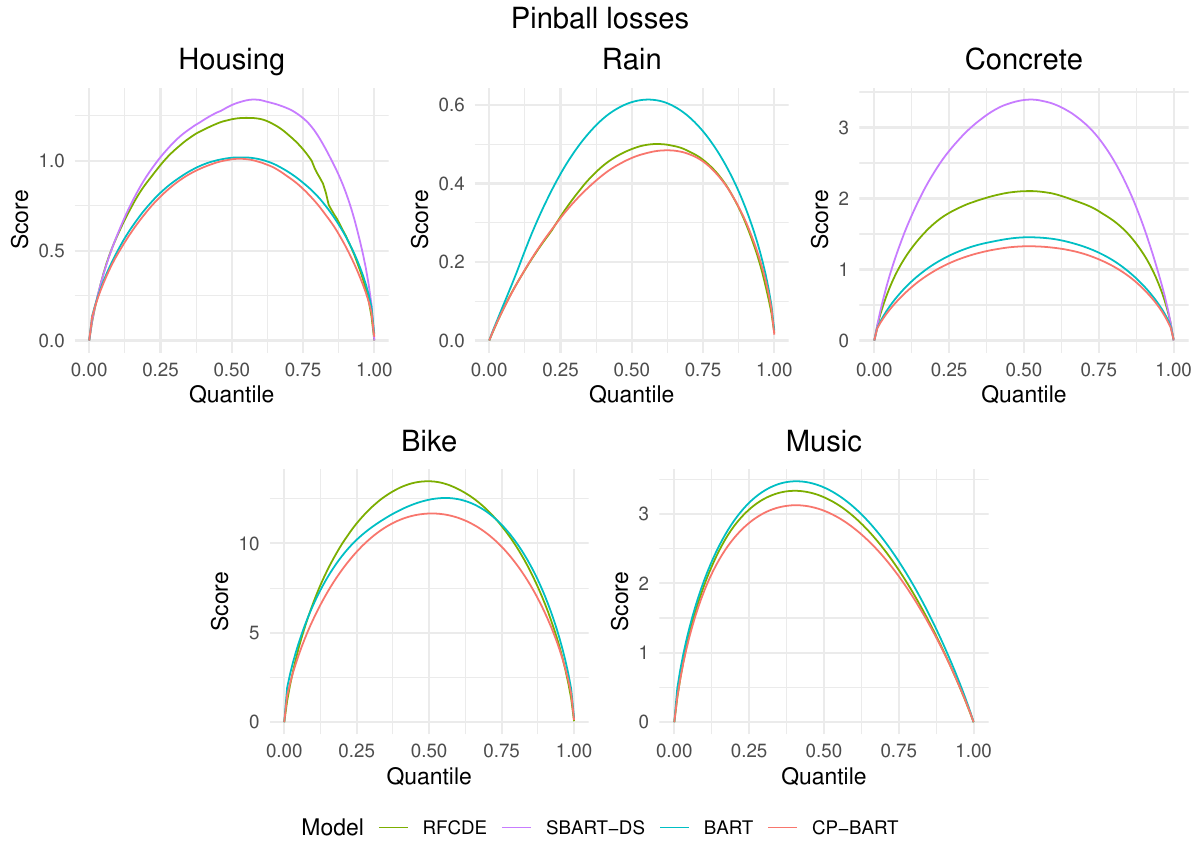}
    \label{fig:pinball}
    \caption{Pinball loss functions $\text{PL}(\alpha)$ plotted against quantile $\alpha$ for the five real-world datasets. The loss functions for the first four datasets are computed using 10-fold cross validation, and for the single pre-specified split for Music, as outlined in the text. Lower values correspond to increased accuracy.}
\end{figure}

\begin{figure}[H]
    \centering
    \includegraphics[width=0.8\linewidth]{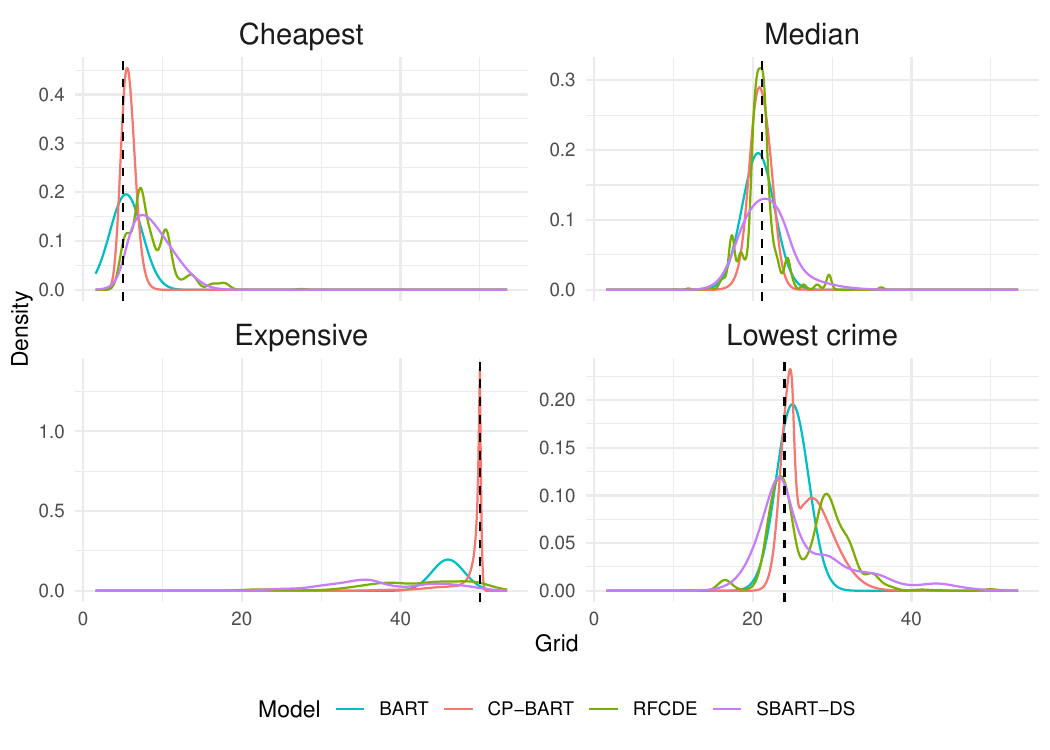}
    \caption{The plot shows the out-of-sample predicted conditional distributions across all methods for four different observations in the Housing dataset. The models were trained on all remaining data. The dashed vertical lines mark the observations.}
    \label{fig:housepred}
\end{figure}

\begin{figure}[tbh]
    \centering
    \includegraphics[width=\linewidth]{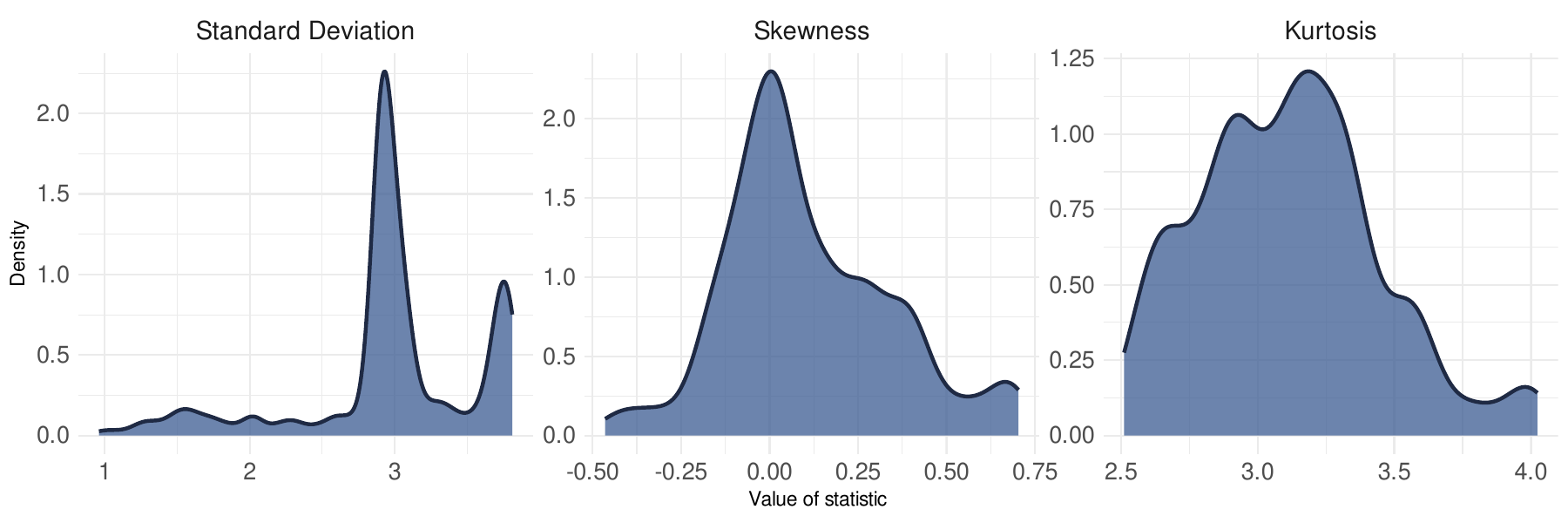}
    \caption{Kernel density estimates of the standard deviation, skewness and kurtosis of the predictive densities for the $n=1,030$ observations in the concrete dataset. Each panel corresponds to one summary statistic; the $x$-axis shows the value of that statistic and the $y$-axis shows the estimated density.}
    \label{fig:momentkde}
\end{figure}

 \subsection{Additional Properties}\label{sec:concreteresults}
 We illustrate some additional properties of CP-BART using the concrete dataset. First, Figure~\ref{fig:momentkde} plots density estimates of the variance, skewness and kurtosis of the predictive densities~\eqref{eq:predp} for the $n$ observations.
 They show that accounting for the covariates through the copula at~\eqref{eq:regcop} can produce a high level of variability in the distribution across observation. Second, to illustrate the impact of the covariates on the dependence structure of the fitted copula process we compute the Spearman correlation between $Y_{n+1}|x_{n+1}$ and $Y_{n+2}|x_{n+2}$, denoted
 as $\rho^S(x_{n+1},x_{n+2})$. 
Figure~\ref{fig:depconcall}(a) plots this for covariate values that are are jointly varied from 0 to 1, 
which are their minimum and maximum values. By construction, $\rho^S=1$ when $x_{n+1}=x_{n+2}$. The correlation 
declines smoothly as 
$|x_{n+1}-x_{n+2}|$ increases to 1, which reflects the smoothness in the tree ensemble. Third, we compare estimation of $\eta$ conditioning on $p_Y$ (the approach used here), with a joint 
posterior that also accounts for uncertainty in $p_Y$ which is modeled using a Bayesian nonparametric infinite mixture. 
Figure~\ref{fig:depconcall}(b) plots the CRPS values for the predictive densities for all $n$ observations, and accounting for uncertainty in $p_Y$ has minimal impact; a result often observed in applied copula modeling. 
Further details and results on the dependence of the copula process and the joint estimation are given in the SM, Section H.

\begin{figure}[tbh]
    \centering
    \includegraphics[width=\linewidth]{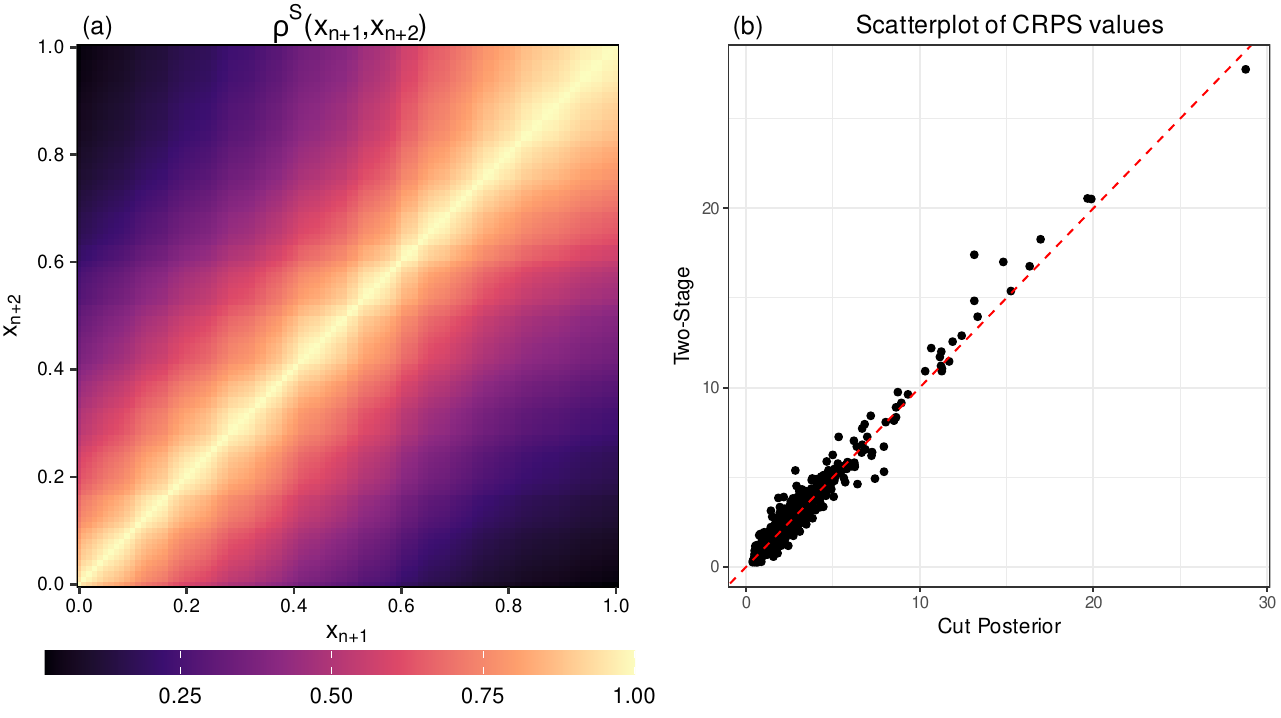}
    \caption{Panel~(a): Spearman's rho values $\rho^S(x_{n+1},x_{n+2})$ from the copula process fitted to the concrete dataset with covariates rescaled to the interval $[0,1]$ when $x_{n+1}$ and $x_{n+2}$ are varied from  0 to 1. Panel~(b): Scatterplot of the CRPS values of the predictive densities for the $n$ observations for the two-stage estimator (vertical axis) and a joint Bayesian estimator of $\eta,p_Y$ (horizontal axis). The mean of the CRPS values over all samples from the joint Bayesian estimator is $1.99$. Further details can be found in the SM, Section H.}
    \label{fig:depconcall}
\end{figure}

\section{Discussion}\label{sec:discussion} 
This paper is the first to show how to construct a Gaussian copula process from a BART model, and how it provides a robust and effective approach to extend BART to distributional regression. To do so, we make the following four contributions.

First, Proposition~\ref{lem:transport} shows that the copula process model is connected to the BART model for the pseudo-responses by an optimal transport map. This map is used to derive computationally efficient expressions for the conditional predictive distribution at~\eqref{eq:predp}, as well as the conditional mean and quantile functions.  
Second, our theoretical results extend those of \citet{rockova19a} who study the posterior of BART in a correctly specified Gaussian regression model with a fixed leaf-node variance.  In contrast, we allow the copula process model to be misspecified and incorporate a hyperprior on the leaf node variance $c$. The KL projection of $c$ depends on the conditional means and conditional variances of the transformed responses, therefore the hyperparameter is necessary to estimate $c$ from the data.We show that the KL projection of the data-generating process onto the model class exists, is unique and has a closed form expression. We show that the fractional posteriors for the conditional densities at the observed covariate values concentrate around this projection   in average Hellinger distance at  rate $n^{-\alpha/(2\alpha+q_0)}(\log n)^{2\alpha/(2\alpha+q_0)}$, thereby adjusting to both, the smoothness of the regression function through $\alpha$ and the number of active covariates $q_0$.
Third, while evaluating the posterior directly is computationally demanding, it can be made tractable by augmenting the likelihood with the leaf node values $\mathcal{M}$. Although we stress that the implicit copula is strictly formed with $\mathcal{M}$ integrated out; its reintroduction here is purely because the resulting augmented posterior can be evaluated using an MCMC sampler that exploits the fast sampling steps in~\cite{ChiGeoMcC2010}. Thus, like the original BART algorithm, CP-BART is scalable with respect to sample size $n$, number of trees $m$ and number of covariates $p$. Fourth, the simulation studies and five real data examples demonstrate scalability, accuracy of the predictive density, and both the accuracy and well-calibrated coverage of the conditional quantile function. 
In these empirical examples, CP-BART dominates the original Gaussian BART model, the distributional random forest method of~\cite{rfcde} and the distributional regression BART model of~\cite{LiLinMur2023}. It provides sharp posterior probability intervals that maintain the desired coverage throughout. 

Extending modern machine learning methods to distributional regression is a topic of substantial interest; for examples, see~\cite{schlosser2019distributional} and \cite{pmlr-v119-duan20a}. The CP-BART model contributes to this literature. It extends and refines the work by~\cite{Klein2019} and~\cite{smithklein21} for a regularized linear regression model to the nonparametric BART model. It also the first paper to lay the theoretical foundations of such implicit Gaussian copula processes. The CP-BART model is also related to the approach of~\cite{kowalwu25}, who propose to efficiently estimate the joint posterior of parametric regressions and unknown transformations using the Bayesian bootstrap. In contrast, the focus here is on forming the copula process model using an optimal transport map, and computing inference using an efficient MCMC sampler for an augmented posterior. Like the recently proposed alternative methods of~\cite{LiLinMur2023} and \cite{OHagRoc2025}, CP-BART demonstrates  the potential of BART not only for point prediction, but also for distributional prediction. 

We finish by highlighting directions for future work. By considering different specifications for the BART model at~\eqref{eq:regression}, different copula processes can be derived. For example,~\cite{UmLinSinBan2023} consider extending BART to allow for skew normal disturbances and an $N$-dimensional multivariate response variable. The former leads to a skew normal copula process, which can capture asymmetric dependence between response values. Such a copula process can be estimated using an extension of our MCMC scheme by exploiting the conditionally Gaussian representation of the skew normal distribution as in~\cite{deng2025large}. The latter extends the copula at~\eqref{eq:gcopprocess} to $Nn$ dimensions that captures both cross-variable and inter-observation dependence. However, this requires specification of a suitable prior on the leaf nodes, which are integrated out to construct the copula process. Finally, more complex BART model specifications also have implicit copula processes, but with more parameters and posteriors that may be more difficult to traverse using MCMC. In this case, exploring the application of variational inference techniques to compute approximate posterior inference at scale may be attractive.
From a theoretical perspective, future work could include extensions to the fractional posteriors in the limiting case $\rho\to 1$, to predictions at new covariate values and to the two-stage estimator with an estimated $F_Y$.

\bibliography{bibliography}

\appendix


\renewcommand\thetheorem{\thesection.\arabic{theorem}} \renewcommand\thelemma{\thesection.\arabic{lemma}} \renewcommand\thedefinition{\thesection.\arabic{definition}} 
\renewcommand\thefigure{\thesection.\arabic{figure}} 
\renewcommand\thetable{\thesection.\arabic{table}}    
\numberwithin{equation}{section}
\numberwithin{figure}{section}
\numberwithin{table}{section}

\end{document}